\documentclass[twocolumn,amsmath,superscriptaddress,amssymb,nopacs,nofootinbib,prx]{revtex4-2}

\usepackage[usenames,dvips]{color}
\usepackage{graphicx}
\usepackage{dcolumn}
\usepackage{bm}
\usepackage{mathtools}
\usepackage{epstopdf}
\usepackage{esint}
\usepackage{xcolor}
\usepackage{dsfont}
\usepackage[outline]{contour}
\usepackage{nicefrac}

\usepackage[colorlinks=true,citecolor=magenta,linkcolor=magenta,urlcolor=magenta]{hyperref}

\newcommand{\bea}{\begin{eqnarray}}
\newcommand{\eea}{\end{eqnarray}}
\newcommand{\bt}{\textbf}

\newcommand{\phd}{\phantom{\dag}}
\newcommand{\ph}{\phantom{.}}

\newcommand{\noi}{\noindent}
\newcommand{\no}{\nonumber}

\newcommand{\PK}[1]{\textcolor{black}{#1}}

\begin{document}
\def\v#1{{\bf #1}}

\title{Superfluid Stiffness and Josephson Quantum Capacitance:\\Adiabatic Approach and Topological Effects}

\author{Jun-Ang Wang}
\email{wangjunang@itp.ac.cn}
\affiliation{CAS Key Laboratory of Theoretical Physics, Institute of Theoretical Physics, Chinese Academy of Sciences, Beijing 100190, China}
\affiliation{School of Physical Sciences, University of Chinese Academy of Sciences, Beijing 100049, China}

\author{Mohamed Assili}
\email{m.assili@itp.ac.cn}
\affiliation{CAS Key Laboratory of Theoretical Physics, Institute of Theoretical Physics, Chinese Academy of Sciences, Beijing 100190, China}

\author{Panagiotis Kotetes}
\email{kotetes@itp.ac.cn}
\affiliation{CAS Key Laboratory of Theoretical Physics, Institute of Theoretical Physics, Chinese Academy of Sciences, Beijing 100190, China}

\vskip 1cm
\begin{abstract}
We bring forward a unified framework for the study of the superfluid stiffness and the quantum capacitance of superconducting platforms exhibiting conventional spin-singlet pairing. We focus on systems which in their normal phase contain topological band touching points or crossings, while in their superconducting regime feature a fully gapped energy spectrum. Our unified description relies on viewing these two types of physical quantities as the charge current and density response coefficients obtained for ``slow" spatiotemporal variations of the superconducting phase. Within our adiabatic formalism, the two coefficients are given in terms of Berry curvatures defined in synthetic spaces. Our work lays the foundations for the systematic description of topological diagonal superfluid responses induced by singularities dictating the synthetic Berry curvatures. We exemplify our approach for concrete one- and two-dimensional models of superconducting topological (semi)me\-tals. We discuss topological phenomena which arise in the superfluid stiffness of bulk systems and the quantum capacitance of Josephson junctions. We show that both coefficients become proportional to a topological invariant which counts the number of topological touchings/crossings of the normal phase band structure. These topological effects can be equivalently viewed as manifestations of chiral anomaly. {\color{black}Our predictions appear experimentally testable in topological semimetals with proximity-induced pairing, such as in graphene-superconductor hybrids at charge neutrality}.
\end{abstract}

\maketitle

\section{Introduction}

It is well-known that the superfluid stiffness of a single-band conventional superconductor (SC) is inversely proportional to the effective mass of this band~\cite{Schrieffer}. This result further predicts that, when this band is nondispersive, i.e., flat, the superfluid stiffness that it carries va\-nishes. Strikingly, recent theo\-re\-ti\-cal~\cite{Torma,PeottaLieb,LongLiang,PeottaTormaBAB,RossiRev,Jonah} and expe\-ri\-mental~\cite{Cao,Yankowitz} works have esta\-bli\-shed that the above prediction breaks down for multiband SCs. Indeed, a careful analysis shows that the superfluid stiffness of a flat band is not only nonzero~\cite{Torma}, but it can be even bounded from below. Such a restriction has been understood using topological arguments~\cite{Ahn,Randeria,Pikulin,Julku,BABbounds,BABbounds2}. Specifically, the explanation for this counter intuitive result relies on the fact that the superfluid stif\-fness of a given band in a multiband SC takes an additional contribution which has purely interband character~\cite{Torma}. This extra contribution is also termed geo\-metric when it happens to be given by the quantum metric of the occupied bands~\cite{Torma}.

The discovery of topological bounds on the superfluid stiffness naturally leads to the following question: is it possible to identify systems whose superfluid stiffness is not only bounded by a topological invariant, but instead, it is equal to a topological invariant itself? This pursuit is crucial, since it paves the way to a quantized superfluid stiffness which can be robust against perturbations. This, in turn, can uncover a plethora of topologically equivalent platforms governed by the same universal superfluid response. Moreover, it can lead to a rich interplay between quantum geo\-metry and topology in superfluid transport.

In this Manuscript, we bring forward that superconducting topological semimetals (STSs) provide a playground for observing the quantization of the total superfluid stiffness due to the nontrivial topology in their normal phase. To transparently present the above result and set the stage for the search of topological superfluid responses in ge\-ne\-ral, we here put forward a new approach for the inve\-stigation of the superfluid stiffness. In particular, we propose to equivalently define the superfluid stiffness tensor elements $D_{ij}=D_{ji}$ as the coefficients which relate the charge current components $J_i(\bm{r})$ to the spatial derivatives of the superconducting phase $\phi(\bm{r})$, through the relation:
\begin{align}
J_i(\bm{r})=-D_{ij}\frac{\partial_j\phi(\bm{r})}{2}
\label{eq:AdiabDefSS}
\end{align}

\noi with $i,j=x,y,z$ for 3D systems. The above relation is expressed in a unit system where the reduced Planck constant $\hbar$ and the electric charge unit $e$ are set to unity.

We focus on metals and semimetals which preserve time-reversal symmetry (TRS), and also experience conventional spin-singlet superconductivity. The combination of these two features ensures that there exists a full gap in their ener\-gy spectrum and, in turn, that the current can be obtained by assu\-ming that the phase varies slowly in space. This alternative, but fully equivalent point of view, lies in the core of the adia\-ba\-tic approach proposed here. \PK{Using our framework, each superfluid stif\-fness element $D_{ij}$ is expressed as a pro\-duct of the normal phase group velocity ope\-ra\-tor $\hat{\upsilon}_i(\bm{p})$ and a Berry curvature operator $\hat{\cal F}_{p_j\phi}(\epsilon,\bm{p},\phi)$ which is defined in a synthetic space spanned by the ener\-gy $\epsilon$, the momentum $p_j$, and the phase $\phi$. Our reformulation reveals that, while the ele\-ments of the superfluid stif\-fness tensor are not in ge\-ne\-ral to\-po\-lo\-gi\-cal invariant quantities, they can still take quantized values for STSs. The quantization in the situa\-tions of interest stems from the presence of monopoles in the synthetic Berry curvature mentioned above. We remark that the quantization discussed here is not universal, i.e., while the superlfuid stiffness is proportional to a to\-po\-lo\-gi\-cal invariant, its ``quantum" is in units which involve a material dependent constant. {\color{black}Our companion work shows that the quantization is robust against weak uncorrelated disorder while, in 2D, it additionally results in a universal topological quantum admittance effect~\cite{WJA}.}}

In what follows, we first motivate our adia\-ba\-tic approach, subsequently formulate it, and finally apply it to a variety of SCs defined in different spatial dimensions. We focus on SCs which in their normal phase can be metallic, or, semimetallic featuring to\-po\-lo\-gi\-cal band touching points (BTPs). Among others, this work provides further support to Ref.~\onlinecite{WJA}, wherein which, the adia\-ba\-tic approach presented here in detail, was first introduced. We remind the reader that in Ref.~\onlinecite{WJA} we provide a trans\-pa\-rent explanation for the predicted quantization of the superfluid stif\-fness in bulk superconducting graphene~\cite{KopninSoninPRL,KopninSoninPRB,KopninLayers,Peltonen}, in terms of the nontrivial topological properties of the STS which stem from the Dirac BTP.

As {\color{black}mentioned} in Ref.~\onlinecite{WJA}, the topological properties of superconducting graphene can be also viewed as the result of 1D chiral anomaly. The emergence of the latter is further clarified in this work. The added benefit of establishing such a connection between chiral anomaly and superfluid response, is that it brings to light ano\-ther quantity which observes topological effects in STSs. This quantity is the {\color{black}quantum capacitance per area $c_{\cal Q}$~\cite{Brey,Ouyang,Jena,GrapheneCQ,Ensslin,Ponomarenko},} that dictates a Josephson junction built from two superconducting plates which are kept at a voltage difference $V$ and are separated by a spacer consisting of {\color{black}a quantum material} and a high-efficiency dielectric.

The standard definition of the here-termed Josephson quantum capacitance (JQC) is given through the relation $\rho_c=c_{\cal Q}V$, where $\rho_c$ denotes the excess charge density appearing on each one of the Josephson junction capacitor plates due to the voltage bias. In analogy to the alternative adiabatic approach that we introduce for the superfluid stiffness, gauge invariance also allows us to bring forward an equi\-va\-lent definition for the JQC, which instead involves the time derivative of the superconducting phase difference characterizing the Josephson junction. With no loss of generality, we assume that the value of the superconducting phase is zero in one of the leads and equal to $\phi$ in the other. Under this gauge choice, we propose to define the JQC through the following expession:
\begin{align}
\rho_c(t)=c_{\cal Q}\frac{\partial_t\phi(t)}{2}
\label{eq:AdiabDefJQC}\,.
\end{align}

Equations~\eqref{eq:AdiabDefSS} and~\eqref{eq:AdiabDefJQC} allow us to unify the superfluid stiffness and JQC in superconducting (semi)metals, by viewing them as the adiabatic charge density and current responses to spatiotemporal gradients of the superconducting phase. This unified picture further hints at a JQC of topological origin. Indeed, such a possibility was first discussed in Ref.~\onlinecite{WJA} for Josephson junctions invol\-ving {\color{black}superconducting (strained) graphene~\cite{Peltonen,BreyFertig,Wakabayashi,covaci,bruno,Roy,V-Kauppila}}. There, it was also shown that the JQC for strained graphene is directly connected to the quantum metric of {\color{black}the zero energy state of the so-called pseudo-Landau le\-vels~\cite{Graphene,NeekCovaciAmal,Neek-Amal,Salerno,Shuze,Nigge,Hsu,Jamotte,Tianyu22}}. Hence, our approach also unveils connections between diagonal (non-Hall type) topological responses and the quantum geometry ari\-sing in currently experimentally accessible materials and nano\-de\-vices.

The presentation of the above topics is organized as follows. First, in Sec.~\ref{sec:StandardSuperStiffness}, we review the standard approach employed for the eva\-lua\-tion of the superfluid stiffness, which we apply in Sec.~\ref{sec:SSTopoMetals} for STSs in various dimensions. In Sec.~\ref{sec:StiffnessReform} we proceed with introducing our alternative method to obtain the superfluid stiffness. We exemplify how our method works in Secs.~\ref{sec:Adiab1D} and~\ref{sec:Adiab2D}, where we focus on the 1D and 2D systems discussed in Sec.~\ref{sec:SSTopoMetals}. Next in line is the discussion of topological effects in the JQC, which are analyzed in Section~\ref{sec:JQC}. There, we present the standard approach to JQC, we introduce the here proposed reformulated method, and afterwards evaluate the JQC for the systems investigated in Secs.~\ref{sec:Adiab1D} and~\ref{sec:Adiab2D}. Section~\ref{sec:Zeeman} considers the effects of a Zeeman field on the quantized phenomena encountered above. Section~\ref{sec:Conclusions} summarizes our findings and provides an outlook. {\color{black}Finally, Appendices~\ref{app:AppendixA}-\ref{app:AppendixE} provide further technical details.}

\section{Standard Theory of Superfluid Transport}\label{sec:StandardSuperStiffness}

In this section, we review the routinely-used procedure to obtain the current and superfluid stiffness which dictate the supefluid transport in a SC with a conventional pairing gap $\Delta\geq0$. In the standard approach, one desires to obtain the electrical current $\bm{J}$ as a response to a spatially uniform and time-independent vector potential $\bm{A}$. The elements of the superfluid stif\-fness tensor $D_{ij}$ are symmetric in $i\leftrightarrow j$, and are defined through the relation:
\begin{align}
D_{ij}=-\frac{1}{2}\left(\frac{\partial J_i}{\partial A_j}+\frac{\partial J_j}{\partial A_i}\right)_{\bm{A}=\bm{0}}\,. 
\end{align}

In the absence of disorder, the SC of interest is described by the following generic bulk Hamiltonian:
\begin{align}
\hat{H}(\bm{p})=\hat{h}(\bm{p})\tau_3+\hat{\Delta}\tau_1.\label{eq:BulkHam}
\end{align}

\noi Here, $\bm{p}$ is the momentum labeling the energy dispersions of the bulk SC. The matrices $\tau_{1,2,3}$ define Pauli matrices acting in Nambu space. The latter space is spanned by electrons with spin up and momentum $\bm{p}$, and, their hole partners related by time reversal (TR) with spin down and momentum $-\bm{p}$. The normal phase Hamiltonian $\hat{h}(\bm{p})$ respects TRS while the pairing matrix $\hat{\Delta}$ is $\bm{p}$-independent and thus symmetric. Although, these two matrices will be specified later on, we here stress that our framework and the results obtained in this work hold for sy\-stems which preserve the full SU(2) spin rotational group or at least a U(1) subgroup of it~\cite{Altland1997,Schnyder2008}. In the latter case odd-under-inversion Rashba-type of spin-orbit coupling terms are also allowed as long as they are orien\-ted in the direction of the spin quantization axis.

To obtain the current $\bm{J}$ using linear response theo\-ry, one needs to evaluate the expectation values of the paramagnetic and diamagnetic current operators $\hat{\bm{J}}^{(p)}$ and $\hat{\bm{J}}^{(d)}$, respectively. These are determined by including in the Hamiltonian the spatially uniform and time-independent vector potential $\bm{A}$. The latter enters through the minimal coupling substitution $\bm{p}\mapsto\bm{p}+\bm{A}\tau_3$. At lowest order in $\bm{A}$, the current operators are determined by the expressions:
\bea
\hat{J}_i^{(p)}(\bm{p})&=&-\hat{\upsilon}_i(\bm{p})\mathds{1}_\tau\,,\\
\hat{J}_i^{(d)}(\bm{p})&=&-\partial_{p_j}\hat{\upsilon}_i(\bm{p})\tau_3A_j\equiv-\partial^2_{p_jp_i}\hat{H}(\bm{p})A_j\,,\label{eq:DiamCurrent}
\eea

\noi where $\hat{\bm{\upsilon}}(\bm{p})=\partial_{\bm{p}}\hat{h}(\bm{p})$ denotes the Bloch electron group ve\-lo\-ci\-ty in the normal phase. Note that the equivalence in the second row of Eq.~\eqref{eq:DiamCurrent} holds only by virtue of the $\bm{p}$-independent pairing gap considered in this work. Moreover, we remark that in the above we adopted the convention of repeated index summation. This is also considered throughout the remainder of this manuscript. In addition, we employed $\mathds{1}_\tau$ to denote the identity matrix in Nambu space. In most instances unit matrices are dropped for notational convenience.

Given the above, we find that the total current per volume flowing in the bulk of the SC reads as:
\bea
J_i&=&-\int dP\ph{\rm Tr}\Big[\hat{\upsilon}_j(\bm{p})\mathds{1}_\tau\hat{G}(\epsilon,\bm{p})\hat{\upsilon}_i(\bm{p})\mathds{1}_\tau\hat{G}(\epsilon,\bm{p})\Big]A_j\no\\
&&-\int dP\ph{\rm Tr}\Big[\hat{G}(\epsilon,\bm{p})\partial_{p_jp_i}^2\hat{H}(\bm{p})\Big]A_j\,,
\eea

\noi where the symbol ``Tr'' denotes trace over all internal degrees of freedom. In addition, we employed for compactness the shorthand notation:
\begin{align}
\int dP\equiv\int_{\rm BZ}\frac{d\bm{p}}{(2\pi)^d}\int_{-\infty}^{+\infty}\frac{d\epsilon}{2\pi}\,.
\end{align}

The momenta are here defined in a $d$-dimensional Brillouin zone (BZ), since the SC is considered to be a crystalline material defined in $d$ spatial dimensions. The conclusions to be obtained also hold when the BZ is replaced by a more general compact space and is crucial for deriving the standard expressions for superfluid stiffness, cf Ref.~\onlinecite{UchoaSeo}. In the above, we also employed $\hat{G}(\epsilon,\bm{p})$, which is the zero temperature Euclidean Green function, whose inverse satisfies $\hat{G}^{-1}(\epsilon,\bm{p})=i\epsilon+B-\hat{H}(\bm{p})$. Here, $B$ is an energy scale which sets the Bogoliubov-Fermi level and arises from the coupling of electrons to a Zeeman field. Moreover, we introduced the energy variable $\epsilon\in(-\infty,+\infty)$ which is obtained as the zero-temperature limit of the fermionic Matsubara frequencies~\cite{AltlandSimons}.

From the above results we immediately obtain the expression for the elements of the superfluid stiffness tensor:
\bea
D_{ij}&=&\int dP\ph{\rm Tr}\Big[\hat{\upsilon}_j(\bm{p})\mathds{1}_\tau\hat{G}(\epsilon,\bm{p})\hat{\upsilon}_i(\bm{p})\mathds{1}_\tau\hat{G}(\epsilon,\bm{p})\Big]\no\\
&+&\int dP\ph{\rm Tr}\Big[\hat{G}(\epsilon,\bm{p})\partial_{p_jp_i}^2\hat{H}(\bm{p})\Big]\,.\label{eq:Dprefinal}
\eea

\noi Elementary manipulations presented in Appendix~\ref{app:AppendixA}, reveal that the diamagnetic contribution is exactly cancelled out by a term contained in the pa\-ra\-mag\-netic contribution~\cite{RossiRev}. As a result, the superfluid stiffness ends up to be given only by the following expression:
\begin{align}
D_{ij}=\int dP\ph{\rm Tr}\Big\{\hat{\upsilon}_j(\bm{p})\tau_3\big[\tau_3,\hat{G}(\epsilon,\bm{p})\big]\hat{\upsilon}_i(\bm{p})\mathds{1}_\tau\hat{G}(\epsilon,\bm{p})\Big\},
\end{align}

\noi where the presence of the commutator $\big[\tau_3,\hat{G}(\epsilon,\bm{p})\big]$ gua\-ran\-tees that the superfluid stiffness is nonzero only for a nonzero pairing gap $\hat{\Delta}$. To make analytical progress, we restrict to the relevant case $\big[\hat{h}(\bm{p}),\hat{\Delta}\big]=\hat{0}$, which yields:
\begin{align}
\hat{G}(\epsilon,\bm{p})=-\frac{i(\epsilon-iB)+\hat{h}(\bm{p})\tau_3+\hat{\Delta}\tau_1}{(\epsilon-iB)^2+\hat{h}^2(\bm{p})+\hat{\Delta}^2},
\label{eq:GreenFunction}
\end{align}

\noi and results in the relation $\tau_3[\tau_3,\hat{G}(\epsilon,\bm{p})]=-\hat{{\cal D}}(\epsilon,\bm{p})\tau_1$, where we introduced the matrix operator:
\begin{align}
\hat{{\cal D}}(\epsilon,\bm{p})
=\frac{2\hat{\Delta}}{(\epsilon-iB)^2+\hat{E}^2(\bm{p})}\,,\no
\end{align}

\noi along with $\hat{E}(\bm{p})=\sqrt{\hat{h}^2(\bm{p})+\hat{\Delta}^2}$. After carrying out the trace in Nambu space, we obtain the expression:
\begin{align}
D_{ij}=\int dP\ph{\rm tr}\Big[\hat{\upsilon}_i(\bm{p})\hat{{\cal D}}(\epsilon,\bm{p})\hat{\upsilon}_j(\bm{p})\hat{{\cal D}}(\epsilon,\bm{p})\Big]\,.\label{eq:StiffnessMain}
\end{align}

\noi The symbol ${\rm tr}$ denotes trace over the degrees of freedom spanning the matrix space in which $\hat{h}(\bm{p})$ is defined.

In certain instances it is more convenient to express the superfluid stiffness as a band property. Such a procedure is presented in Appendix~\ref{app:AppendixB}, and allows us to link our results with previous works~\cite{Torma,PeottaLieb,LongLiang,PeottaTormaBAB,RossiRev}. Notably, however, our for\-ma\-lism presents a new feature. This is the inclusion of the Zeeman coupling to a magnetic field which sets the Fermi level of the Bogo\-liu\-bov energy bands and, thus, controls the occupancy of each energy dispersion.

In experiments, the application of an external Zeeman field in principle allows to isolate the here-sought-after quantized contributions to the superfluid stiffness~\cite{WJA}, which stem from topological BTPs of the nonsuperconducting Hamiltonian. As mentioned earlier, we restrict to SCs which preserve the full SU(2) spin rotational invariance or a U(1) subgroup of it. In the former case, there is in principle no restriction on the direction of the applied Zeeman field, other than being such so that it can sustain superconductivity. In contrast, in the second case, the field is additionally required to be oriented along the spin axis which ge\-ne\-ra\-tes the U(1) subgroup.

\section{Superfluid Stiffness of Superconducting Topological Metals}\label{sec:SSTopoMetals}

The expression in Eq.~\eqref{eq:StiffnessMain} is particularly convenient to use when examining the properties of SCs which in their nonsuperconducting phase are either to\-po\-lo\-gi\-cal se\-mi\-me\-tals containing BTPs or metals exhibiting topologically protected dispersive electrons. To show this, we first assume that $\hat{h}(\bm{p})$ satisfies the property $\hat{h}^2(\bm{p})=\varepsilon^2(\bm{p})\mathds{1}_h$, where $\mathds{1}_h$ is the identity matrix in the matrix space in which $\hat{h}(\bm{p})$ is defined. Under this condition, the energy integration in Eq.~\eqref{eq:StiffnessMain} is straightforward and by further considering $B=0$ and assuming $\hat{\Delta}=\Delta\mathds{1}_h$, it leads to:
\begin{align}
D_{ij}=\int_{\rm BZ}\frac{d\bm{p}}{(2\pi)^d}\frac{\Delta^2}{E^3(\bm{p})}\,{\rm tr}\big[\hat{\upsilon}_i(\bm{p})\hat{\upsilon}_j(\bm{p})\big]\,.\label{eq:StiffnessMainSemi}
\end{align}

Apart from the trivial case in which $\hat{h}(\bm{p})$ is simply given by a zero-dimensional matrix, i.e., the energy di\-sper\-sion $\varepsilon(\bm{p})$ itself, the property $\hat{h}^2(\bm{p})=\varepsilon^2(\bm{p})\mathds{1}_h$ is ty\-pi\-cal for Hamiltonians defined using Clifford algebras. In such situations, the normal phase Hamiltonian can be expressed according to $\hat{h}(\bm{p})=\bm{d}(\bm{p})\cdot\hat{\bm{\Gamma}}$, with the matrices $\big\{\hat{\Gamma}_a,\hat{\Gamma}_b\big\}=2\delta_{ab}\mathds{1}_h$ generating a special orthogonal Clifford algebra. The vector $\bm{d}(\bm{p})$ is expressed as $\bm{d}(\bm{p})=\varepsilon(\bm{p})\bm{n}(\bm{p})$, with the unit vector $\bm{n}(\bm{p})$ being defined in the respective internal space. For such Clifford systems, the superfluid stiffness takes the simplified form:
\begin{align}
D_{ij}=d_h\int_{\rm BZ}\frac{d\bm{p}}{(2\pi)^d}\frac{\Delta^2}{E^3(\bm{p})}\left[\partial_{p_i}\bm{d}(\bm{p})\cdot\partial_{p_j}\bm{d}(\bm{p})\right],\label{eq:StiffnessMainSemiClifford}
\end{align}

\noi where $d_h$ is equal to the matrix dimension of $\hat{h}(\bm{p})$. We note that one can further express the inner pro\-duct appearing inside the brackets according to $\partial_{p_i}\bm{d}(\bm{p})\cdot\partial_{p_j}\bm{d}(\bm{p})=\upsilon_i(\bm{p})\upsilon_j(\bm{p})+\varepsilon^2(\bm{p})\partial_{p_i}\bm{n}(\bm{p})\cdot\partial_{p_j}\bm{n}(\bm{p})$, where we introduced the Bloch group velocity vector $\bm{\upsilon}(\bm{p})=\partial_{\bm{p}}\varepsilon(\bm{p})$. As it has been shown in Ref.~\onlinecite{ChenWei}, the second term has a quantum geometric character, since it is proportional to the elements of the quantum metric tensor $g_{ij}(\bm{p})$ of the occupied bands.

In the remainder of this section, we employ the above results to infer the superfluid stiffness for concrete experimentally-accesible SCs. Specifically, we focus on continuum models which describe systems con\-tai\-ning topological band crossings and band touchings. For the moment, we restrict ourselves to inferring only the contribution of these regions of momentum space in the band structure. Note that this may appear to contradict the assumptions under which Eq.~\eqref{eq:StiffnessMain} was derived since, for this equation to hold, the momenta should be embedded in a compact space. The approach followed here should be understood as aiming at identifying the contribution of the topological band touching and crossing electrons, hence discar\-ding the contributions of electrons from momenta belonging to the remainder of the momentum space, with the latter still assumed to be compact.

\subsection{1D Superconducting Topological Semimetals}\label{sec:TIedge}

\begin{figure}[t!]
\begin{center}
\includegraphics[width=1\columnwidth]{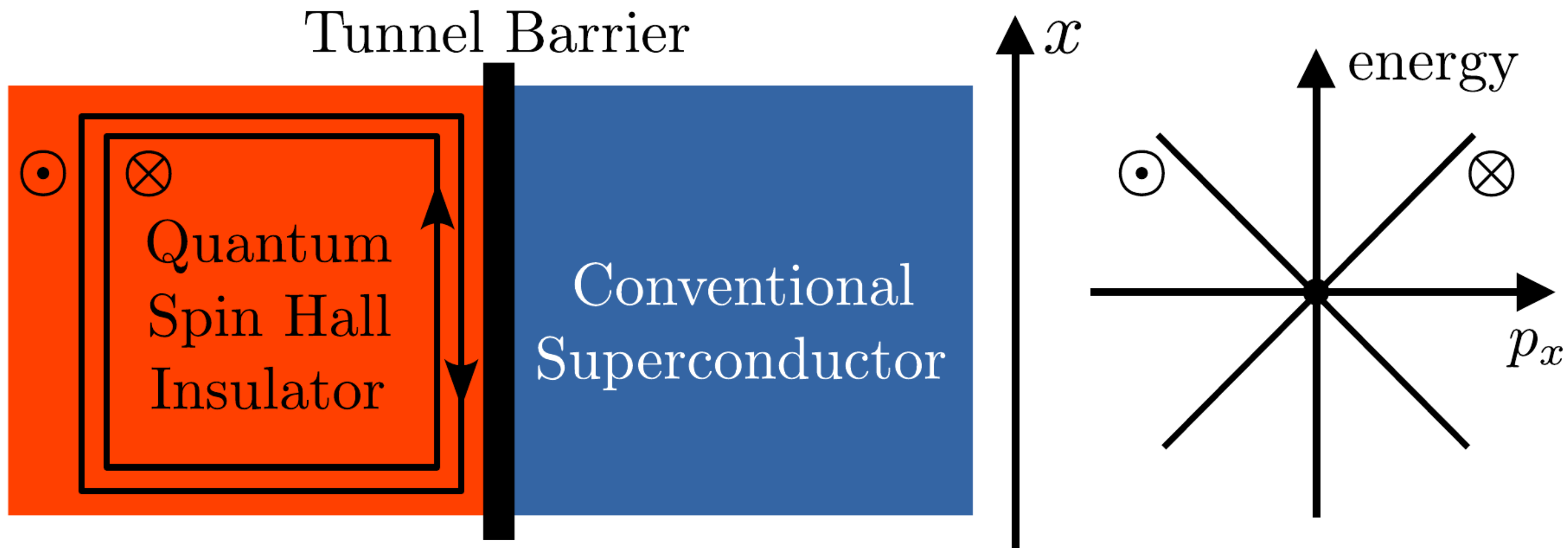}
\end{center}
\caption{Interface of a quantum spin Hall insulator and a conventional spin-singlet superconductor. The spin-filtered helical edge modes appearing on the interface with the superconductor inherit a pairing gap whose strength is controlled by a tunnel barrier. The helical edge modes see opposite spin polarizations denoted $\{\otimes,\odot\}$. Moreover, for low energies, the dispersions of the helical edge modes are linear and feature group velocities $\pm\upsilon_D$. The contribution of the interface helical modes to the superfluid stiffness for supercurrent flow in the $x$ direction is quantized according to $\upsilon_D/\pi$. This quantized contribution can be experimentally disentangled by compa\-ring the superfluid stiffness of the superconducting platform in the topologically trivial and nontrivial phases of the quantum spin Hall insulator, in which edge modes are absent and present, respectively.}
\label{fig:PRRFigure1}
\end{figure}

Our first case study concerns strictly 1D SCs and, in particular, the edge of a two-dimensional quantum spin Hall insulator~\cite{KaneMele1,KaneMele2,BHZ,HgTe} which is here assumed to feature a conventional pai\-ring gap due to its proximity to a neighboring bulk SC. See Fig.~\ref{fig:PRRFigure1} for an illustration. In the normal phase, the topological edge harbors a helical electron branch con\-si\-sting of two sub-branches with dispersions {\color{black}$\pm\upsilon_D p_x$} and opposite spin projection. Such a helical branch is described by the normal phase Hamiltonian $\hat{h}_N(p_x)=\upsilon_Dp_xs_z$, with $s_z$ denoting the 3rd spin Pauli matrix and $p_x\in(-p_c,+p_c)$, where $p_c$ a momentum cutoff. Note that the Hamiltonian $\hat{h}_N(p_x)$ is expressed in the electron instead of the Nambu basis. This basis is spanned by the fol\-lo\-wing two electron states $\{|e,\uparrow,\,p_x>,|e,\downarrow,\,p_x>\}$.

To describe the superconductive case, we include the hole excitations and obtain a Hamiltonian $\hat{H}(p_x)$ of the form discussed in Eq.~\eqref{eq:BulkHam}. This Hamiltonian is defined in the Nambu basis $\{|e,\uparrow,\,p_x>,|h,\downarrow,\,-p_x>\}$ which is also introduced by employing the two-component spinor:
\begin{align}
\bm{\Psi}^\dag(p_x)=\Big(\psi_\uparrow^\dag(p_x),\,\psi_\downarrow^\dag(-p_x)\Big).\label{eq:TIspinor}
\end{align}

\noi In the Nambu basis, the superconducting topological edge is described by a Hamiltonian $\hat{h}(p_x)=\upsilon_Dp_x$, since spin is already absorbed in the definition of the basis.

\subsubsection*{Quantization of Superfluid Stiffness and Hints of Topology}

In order to study the superfluid transport for this sy\-stem when $B=0$,\footnote{As we have previously announced, modifying the level occupancy by means of sweeping the strength of an externally imposed Zeeman field (along the spin $z$-axis here), can further disentangle the contribution of topological band crossings and tou\-chings.} we first start from Eq.~\eqref{eq:StiffnessMainSemi} and eva\-lua\-te the superfluid stiffness $D_{xx}$. The algebraic ma\-ni\-pu\-la\-tions are elementary and we end up with:
\begin{align}
D_{xx}=\frac{\upsilon_D}{\pi}\int_{-p_c}^{+p_c}dp_x\,\frac{\upsilon_D\Delta^2}{2E^3(p_x)},
\label{eq:TIedge}
\end{align}

\noi where $E(p_x)=\sqrt{(\upsilon_Dp_x)^2+\Delta^2}$. The integral retains contributions mainly from the neighbourhood of $p_x=0$. This allows us to take the limit $p_c\rightarrow\infty$ and find:
\begin{align}
\int_{-\infty}^{+\infty}dp_x\,\frac{\upsilon_D\Delta^2}{2E^3(p_x)}=1,
\end{align}

\noi which implies that the superfluid stiffness is quantized in units of $\upsilon_D/\pi$. We remark that, while $\upsilon_D$ is material dependent, this behavior can be still viewed as universal in a certain sense. This is because $D_{xx}$ is proportional to a to\-po\-lo\-gical invariant quantity which counts the number of Dirac points in the normal phase Hamiltonian $\hat{h}(p_x)$.

Notably, the above integral also appears in the theory of chiral anomaly for 1D Dirac electrons~\cite{Semenoff,CA,VolovikBook,Kotetes}, and dictates the topological response induced by spatiotemporal variations in the phase of the Dirac mass~\cite{GW}. Here, it is the pairing gap $\Delta$ that plays the role of the Dirac mass, and the phase involved is the supercon\-duc\-ting phase $\phi$, which enters through $\Delta\mapsto \Delta e^{i\phi}$. This connection becomes clear by equivalently viewing the uniform probe vector potential $A_x$ as the result of a constant spatial gradient of the superconducting phase, i.e., $A_x=\partial_x\phi/2$.

The above replacement leads to the following relation:
\begin{align}
J_x=-\upsilon_D\frac{\partial_x\phi}{2\pi}\label{eq:Current1D}
\end{align}

\noi which establishes the connection between the superfluid response and the Goldstone-Wilczek formula~\cite{GW}, albeit the following differences: (i) here it is the charge current $J_x$ instead of the charge density $\rho_c$ which is induced by a spatial gradient of the phase of the Dirac mass, and (ii) an extra factor of $\upsilon_D$ appears due to the exchanged role of charge density and current. In the upcoming sections we clarify how chiral anomaly emerges and explain the different roles played by the physical quantities involved.

\subsection{2D Superconducting Topological Semimetals}\label{eq:2dDiracCone}

After exemplifying our approach for a 1D supercon\-duc\-ting topological band crossing point, we now employ Eq.~\eqref{eq:StiffnessMainSemiClifford} to obtain the superfluid stif\-fness of a single superconducting Dirac cone defined in 2D. This allows to establish a connection to the result found in Ref.~\onlinecite{KopninSoninPRL} for 2D superconduc\-ting monolayer graphene in the Dirac-cone regime, and that was recently also stu\-died in our related work in Ref.~\onlinecite{WJA}. In the Dirac regime and at charge neutrality, the normal phase graphene Hamiltonian consists of two blocks, each of which describes a single valley labelled by $\lambda=\pm1$~\cite{Graphene}. In the Nambu basis, we find:
\begin{align}
\hat{h}_\lambda(\bm{p})=\upsilon_D\big(p_x\sigma_1+\lambda p_y\sigma_2\big)\no
\end{align}

\noi where $\upsilon_D$ is the {\color{black}Dirac} velocity and $\sigma_{1,2,3}$ denote Pauli matrices acting in the sublattice space spanned by the two interpenetrating triangular lattices of graphene~\cite{Graphene}. The valley Hamil\-to\-nians feature identical eigenenergies with $\varepsilon_{\sigma}(\bm{p})=\sigma\varepsilon(\bm{p})$, where $\varepsilon(\bm{p})=\upsilon_D|\bm{p}|$, and $\sigma=\pm1$.

\subsubsection*{Quantization of Superfluid Stiffness}

We now obtain the superfluid stiffness elements for a single Dirac cone appearing in graphene using Eq.~\eqref{eq:StiffnessMainSemiClifford}. Specifically, we focus on $\lambda=1$ and restrict to $B=0$. The symmetry properties which dictate the Hamiltonian of a given graphene Dirac cone imply that $D_{xx}=D_{yy}\equiv D$ and $D_{xy}=0$. We use the relation $\partial_{p_x}\hat{h}_\lambda(\bm{p})=\upsilon_D\sigma_1$, along with the property $\hat{h}_\lambda^2(\bm{p})\propto\mathds{1}_\sigma$ which holds at charge neu\-tra\-lity, and find that both valleys contribute equally to the stiffness, with a single-valley contribution:\footnote{Note also that for graphene, electrons couple to holes of different valleys~\cite{Carlo}. Hence, here the BdG spinor follows from Eq.~\eqref{eq:TIspinor}, but after been suitably adjusted along the lines of Ref.~\onlinecite{Carlo}, in order to also encode the valley degree of freedom.}
\begin{align}
D=2\int\frac{d\bm{p}}{(2\pi)^2}\,\frac{\upsilon_D^2\Delta^2}{E^3(\bm{p})},
\label{eq:DconeChiralAnomaly}
\end{align}

\noi where $E(\bm{p})=\sqrt{\varepsilon^2(\bm{p})+\Delta^2}$ and $(p_x,p_y)\in(-p_c,+p_c)^2$.

Since, similar to the 1D case, also here the integral retains contributions mainly from the neighbourhood of $\bm{p}=\bm{0}$, we extend the integration domain to $\mathbb{R}^2$. We subsequently employ cylindrical coordinates, carry out the trivial integration over the angle in momentum space, and conclude with the expression for a single Dirac cone:
\begin{align}
D=\frac{\Delta}{\pi}\int_0^\infty d\zeta\ph\frac{\zeta}{\sqrt{1+\zeta^2}^3}=\frac{\Delta}{\pi},
\end{align}

\noi where we set $\zeta=\upsilon_D p_x/\Delta$. We therefore recover the result that was first obtained in Refs.~\onlinecite{KopninSoninPRL} and~\onlinecite{KopninSoninPRB}. {\color{black}We remind the reader once again that the above holds at charge neutrality and, thus, a zero chemical potential.} Expressing the respective current $J_x$ in terms of $\partial_x\phi$ yields:
\begin{align}
J_x=-\Delta\frac{\partial_x\phi}{2\pi}\,.
\end{align}

\noi and further coincides with the obtained by Titov and Beenakker for short graphene Josephson junctions~\cite{Titov}, when the limit of small phase differences is considered.

As we demonstrate in the upcoming paragraphs, the quantized contribution of the Dirac cone part of the band structure to the superfluid stiffness for superconduc\-ting graphene can be understood by either exten\-ding the conclusions relating to the emergence of 1D chiral anomaly, or, by directly accounting for the topological properties of the 2D graphene Hamiltonian.

\subsection{Superfluid Stiffness from Higher-Order Band Touching Points in 2D}

A natural extension of a single Dirac cone in 2D is to consider a BTP which features a topological charge of higher order. To model such a situation, we consider the normal phase Hamiltonian in the Nambu basis:
\begin{align}
\hat{h}^{(s)}(\bm{p})=\varepsilon_D\left(\frac{p}{p_D}\right)^{|s|}\Big\{\cos[s\theta(\bm{p})]\sigma_1+\sin[s\theta(\bm{p})]\sigma_2\Big\}\,,\label{eq:HOTBTP}
\end{align}

\noi that gives rise to a single BTP which carries a topo\-lo\-gi\-cal charge of $s\in\mathbb{Z}$ units. In analogy to graphene, the assumption of TRS generally requires additional BTPs to be present. Here, we are inte\-re\-sted in the contribution of only a single BTP described by the Hamiltonian above. In Eq.~\eqref{eq:HOTBTP}, we introduced the angle $\tan[\theta(\bm{p})]=p_y/p_x$, while $\varepsilon_D$ is a characteristic energy scale, and $p_D$ a momentum. As in the previous section, $(p_x,p_y)\in(-p_c,+p_c)^2$, and $p_c$ will be taken to infinity.

We note that the arising rotational symmetry of the given model guarantees that $D_{xx}=D_{yy}\equiv D$ while $D_{xy}=0$. The diagonal elements of the superfluid stif\-fness are more conveniently eva\-lua\-ted using the for\-ma\-lism introduced in Sec.~\ref{sec:SSTopoMetals}. Hence, we follow this section and introduce the two-component unit vector $\bm{n}(\bm{p})=(\cos[s\theta(\bm{p})],\sin[s\theta(\bm{p})])$ along with the energy dispersion $\varepsilon(\bm{p})=\varepsilon_D(p/p_D)^{|s|}$ with $p=|\bm{p}|$. The above steps lead to the expression:
\begin{align}
D^{(s)}=\int\frac{d\bm{p}}{(2\pi)^2}\frac{\Delta^2}{E^3(\bm{p})}\sum_i^{x,y}\Big\{\upsilon_i^2(\bm{p})+\big[\varepsilon(\bm{p})\partial_{p_i}\bm{n}(\bm{p})\big]^2\Big\}.
\end{align}

It is straightforward to confirm that each one of the two terms in the brackets contributes equally to the stiffness. Specifically, the contribution of each term is $[s\varepsilon(p)/p]^2$. By plugging the latter in the expression for the superfluid stiffness yields that:
\begin{align}
D^{(s)}=|s|\frac{\Delta}{\pi}.\label{eq:HOTBTPstiffness}
\end{align}

\noi Hence, we conclude that the absolute value of the topological charge of a BTP is imprinted in its contribution to the superfluid stiffness, which is a property that can be in principle harnessed for its detection.

\subsection{3D Superconducting Topological Semimetals}\label{sec:3DSTS}

Our investigation of topological semimetals concludes with the study of the superfluid stiffness of a single superconducting Weyl cone, which is described by the Nambu-space normal phase Hamiltonian: $\hat{h}(\bm{p})=\upsilon_D\bm{p}\cdot\bm{\sigma}$. As it follows from the analysis of the previous paragraphs, the superfluid stiffness tensor is diagonal and isotropic, i.e., $D_{yz,zx,xy}=0$ and $D_{xx,yy,zz}=D$. We thus obtain:
\begin{align}
D=2\int\frac{d\bm{p}}{(2\pi)^3}\ph\frac{\upsilon_D^2\Delta^2}{E^3(\bm{p})}\simeq\upsilon_D\left(\frac{\Delta}{\pi\upsilon_D}\right)^2\ln\left(\frac{2\Lambda}{{\rm e}\Delta}\right)\,,
\label{eq:WeylconeChiralAnomaly}
\end{align}

\noi where $\Lambda$ corresponds to an ultraviolet energy cutoff and e is the Euler number. This is an approximate result obtained in the limit $\Lambda\gg\Delta$. The above reveals that the outcome in 3D is not independent from the theory's cutoff and, most importantly, it receives negligible contributions from the Weyl point located at $p=0$.

At this point, it is interesting to comment on the dependence of the superfluid stiffness on the superconduc\-ting gap. For a conventional 3D SC, one expects to find that the superfluid stiffness is at least proportional to $\Delta^2$, si\-mi\-lar\-ly to what has been found in the Weyl case. However, the 1D and 2D cases clearly deviate from this standard behavior, thus hinting that different me\-cha\-nisms are responsible for the superfluid transport. The independence of the current on the Dirac mass is typical for chiral anomaly and this is exactly what we observe in 1D. The 2D case appears to constitute the intermediate regime where the pairing gap influences transport but with a scaling that does not follow the usual rule.

Based on the above, we thus conclude that the topological features of Weyl band touching points cannot be discerned in measurements of the superfluid stiffness. Nevetheless, it may still be possible to obtain signatures in the superfluid stiffness when additional external fields are imposed, which lead to higher order current responses. Indeed, the strong topology of 3D systems ty\-pi\-cal\-ly manifests itself in current responses which require the simultaneous presence of two external fields. For instance, this is the case for 3D chiral anomaly in Weyl systems~\cite{Semenoff,CA,Schnyder2008,Burkov}. However, such possibilities go beyond the scope of this work and we plan to address such 3D scenarios in a separate dedicated future work.

\section{Adiabatic Reformulation of Superfluid Transport}\label{sec:StiffnessReform}

The above results point toward the involvement of anomalies and nontrivial topology. As we show below, the underlying role of such phenomena becomes transparent by following an alternative route to evaluate the superfluid stiffness. Specifically, for this purpose, we propose to reformulate the theory for superfluid transport by equivalently considering linear response to the spatial derivatives of the super\-con\-ducting phase. Since a SC is a charged superfluid, the coordinate space gradients $\bm{\nabla}\phi(\bm{r})$ of the superconducting phase $\phi(\bm{r})$ effectively act as a vector potential $\bm{A}(\bm{r})$, since gauge inva\-rian\-ce implies the substitution $\bm{A}(\bm{r})\mapsto\bm{A}(\bm{r})+\bm{\nabla}\phi(\bm{r})/2$.

As it was already mentioned in our introduction, within the here proposed approach, we employ an alternative definition for the superfluid stif\-fness tensor, which is obtained by relating the $i$-th component of the electrical current $J_i(\bm{r})$ to the $j$-th spatial gra\-dient $\partial_j\phi(\bm{r})$ of the phase superconducting phase. Notably, while the standard theory for the superfluid stif\-fness is obtained as a response to a spatially uniform and time-independent vector potential, the reformulation presented here relies on the response to a spatially-varying and time-independent phase bias. Therefore, it is here vital to employ a coordinate space description that properly embodies the nontrivial spatial dependence of the phase. For this purpose, we consider the coordinate-space defined Hamiltonian:
\begin{align}
\hat{H}(\hat{\bm{p}},\bm{r})=\hat{h}(\hat{\bm{p}})\tau_3+\hat{\Delta}\tau_1e^{-i\phi(\bm{r})\tau_3},\label{eq:HamCoordinate}
\end{align}

\noi where now $\hat{h}(\hat{\bm{p}})$ depends on the momentum ope\-ra\-tor which takes the differential form $\hat{\bm{p}}=-i\bm{\nabla}$. Note that within the adiabatic approach, the only restriction on the pairing gap matrix $\hat{\Delta}$ is for it to lead to a fully gapped spectrum for $\phi(\bm{r})=0$. For a uniform $\phi$, the Hamiltonian in Eq.~\eqref{eq:HamCoordinate} respects translational invariance in all directions, since the nonpairing part $\hat{h}(\hat{\bm{p}})$ is assumed to depend only on the momentum ope\-ra\-tor $\hat{\bm{p}}$, and the Hamiltonian coincides with the one in Eq.~\eqref{eq:BulkHam}.

To obtain the total current $\bm{J}$, it is sufficient to evaluate the expectation value of the paramagnetic current operator $\hat{\bm{J}}^{(p)}$. To justify this, we discuss the general expression of the energy functional $E(\bm{r})$ for the gauge invariant vector potential, which is obtained after integrating out the fermions of the SC. Since the SC is assumed to respect TRS for $\bm{A}(\bm{r})=\bm{\nabla}\phi(\bm{r})=\bm{0}$, we obtain that:
\begin{align}
E(\bm{r})=D_{ij}\big[A_i(\bm{r})+\partial_i\phi(\bm{r})/2\big]\big[A_j(\bm{r})+\partial_j\phi(\bm{r})/2\big]/2\,. 
\end{align}

\noi As a result, the electrical current is given by:
\begin{align}
J_i(\bm{r})=-\frac{\delta E(\bm{r})}{\delta A_i(\bm{r})}=-D_{ij}\big[A_j(\bm{r})+\partial_j\phi(\bm{r})/2\big]\,,
\end{align}

\noi which implies that the elements $D_{ij}$ are obtainable from a correlation function with vertices involving the vector potential and the gradient of the supercoducting phase, where each one of these is considered at first order. Hence, since $\bm{A}$ enters at first order, only the paramagnetic current is required to be evaluated, and similarly to the previous section, the total current operator within the present framework is given in terms of the limit:
\begin{align}
\hat{\bm{J}}=-\lim_{\bm{A}\rightarrow\bm{0}}\big[\partial\hat{h}(\hat{\bm{p}}+\bm{A}\tau_3)/\partial \bm{A}\big]\tau_3.
\end{align}

To proceed, we first consider small deviations of the superconducting phase $\phi(\bm{r})$ away from the TR-invariant value $\phi=0$. This allows us to ap\-pro\-xi\-ma\-te the Hamiltonian in the following fashion:
\begin{align}
\hat{H}(\hat{\bm{p}},\bm{r})\approx\hat{h}(\hat{\bm{p}})\tau_3+\hat{\Delta}\tau_1-\hat{\Delta}\phi(\bm{r})\tau_2.
\end{align}

\noi To carry out the linear response program, it is more convenient to expand the superconducting phase in terms of Fourier components $\phi(\bm{q})=\int d\bm{r}\, e^{-i\bm{q}\cdot\bm{r}}\phi(\bm{r})$. To obtain the desired expectation value for the current operator, we employ the zero-temperature Green function method as in the previous section. However, in the present case translational invariance is broken, and the Green function can be either described in coordinate space using two position arguments, or, in momentum space using two momentum arguments. For details see Appendix~\ref{app:AppendixC}.

Starting from the Dyson equation, we take into account the first order correction to the single-particle matrix Green function $\hat{G}$ due to the perturbation term $-\hat{\Delta}\phi(\bm{r})\tau_2$. Specifically, we consider a symmetrized expression of the ensuing Dyson equation, which leads to:
\begin{align}
\hat{G}^{(1)}(\epsilon,\bm{p},\bm{q})\approx\frac{\hat{G}(\epsilon,\bm{p})}{2}\left[(2\pi)^d\delta(\bm{q})-\hat{\Delta}\phi(\bm{q})\tau_2\hat{G}(\epsilon,\bm{p}-\bm{q})\right]\no\\
\quad\qquad\qquad+\left[(2\pi)^d\delta(\bm{q})-\hat{G}(\epsilon,\bm{p}+\bm{q})\hat{\Delta}\phi(\bm{q})\tau_2\right]\frac{\hat{G}(\epsilon,\bm{p})}{2},
\label{eq:modGreenMomentum}
\end{align}

\noi where one observes the involvement of the bare matrix Green function defined in Eq.~\eqref{eq:GreenFunction}. In view of the here-assumed slow spatial variation of $\phi(\bm{r})$, we take the limit $\bm{q}\rightarrow\bm{0}$ and consider a uniform phase gradient $\bm{\nabla}\phi(\bm{r})$, so that $\phi(\bm{r})\approx\bm{\nabla}\phi\cdot\bm{r}$. In this limit, we obtain the translationally-invariant modified matrix Green function which is defined as $\hat{G}^{(1)}(\epsilon,\bm{p})=\int\frac{d\bm{q}}{(2\pi)^d}\, e^{i\bm{q}\cdot\bm{r}}\hat{G}^{(1)}(\epsilon,\bm{p},\bm{q})$, and in the present case takes the form:
\bea
\hat{G}^{(1)}(\epsilon,\bm{p})&\approx&\hat{G}(\epsilon,\bm{p})+\bm{\nabla}\phi\cdot\hat{G}(\epsilon,\bm{p})\frac{\hat{\Delta}\tau_2}{2i}\partial_{\bm{p}}\hat{G}(\epsilon,\bm{p})\no\\
&&-\bm{\nabla}\phi\cdot\big[\partial_{\bm{p}}\hat{G}(\epsilon,\bm{p})\big]\frac{\hat{\Delta}\tau_2}{2i}\hat{G}(\epsilon,\bm{p})\,.
\eea

Having identified the perturbed Green function, we now move ahead and obtain the expectation value for the current, which is given by the following expression:
\begin{align}
\bm{J}=-\int dP\ph{\rm Tr}\left[\hat{\bm{\upsilon}}(\bm{p})\mathds{1}_\tau\hat{G}^{(1)}(\epsilon,\bm{p})\right].
\end{align}

\noi We note that no current flows in the absence of the superconducting phase gradient, since the system preserves TRS. Hence, under the assumed spatial uniformity of the phase gradient $\bm{\nabla}\phi$, the above considerations lead to the following expression for the current:
\bea
&&J_i=-\partial_j\phi\int dP\ph{\rm Tr}\Big\{
\hat{\upsilon}_i(\bm{p})\mathds{1}_\tau\Big[\hat{G}(\epsilon,\bm{p})\hat{\Delta}\tau_2\partial_{p_j}\hat{G}(\epsilon,\bm{p})\no\\
&&\qquad\qquad\qquad\quad\qquad-\big[\partial_{p_j}\hat{G}(\epsilon,\bm{p})\big]\hat{\Delta}\tau_2\hat{G}(\epsilon,\bm{p})\Big]\Big\}/2i.\no
\eea

\noi For a sufficiently weak $\bm{\nabla}\phi$, we can rewrite the term in brackets in the follo\-wing approximate fashion:
\begin{align}
\Big\{\hat{G}(\epsilon,\bm{p})\hat{\Delta}\tau_2\partial_{p_j}\hat{G}(\epsilon,\bm{p})-\big[\partial_{p_j}\hat{G}(\epsilon,\bm{p})\big]\hat{\Delta}\tau_2\hat{G}(\epsilon,\bm{p})\Big\}/2i\no\\\approx
e^{-i\phi\tau_3/2}\hat{\cal F}_{p_j\phi}(\epsilon,\bm{p},\phi)e^{i\phi\tau_3/2}\no
\end{align}

\noi where we introduced the matrix function $\hat{\cal F}_{p_j\phi}(\epsilon,\bm{p},\phi)$, defined as:
\begin{align}
\hat{\cal F}_{p_j\phi}=\nicefrac{1}{2}\big(\partial_\epsilon\hat{{\cal G}}^{-1}\big)\hat{{\cal G}}\big(\partial_\phi\hat{{\cal G}}^{-1}\big)\hat{{\cal G}}\big(\partial_{p_j}\hat{{\cal G}}^{-1}\big)\hat{{\cal G}}-\partial_\phi\leftrightarrow\partial_{p_j}\,.
\end{align}

\noi In the above, we suppressed the arguments of the va\-rious functions for notational convenience and, most importantly, we introduced the matrix Green function through:
\begin{align}
\hat{{\cal G}}^{-1}(\epsilon,\bm{p},\phi)=i\epsilon+B-\hat{{\cal H}}(\bm{p},\phi)\,,\label{eq:SynthGreen}
\end{align}

\noi which is defined in the synthetic energy-momentum-phase space and results from the adiabatic Hamiltonian:
\begin{align}
\hat{{\cal H}}(\bm{p},\phi)=\hat{h}(\bm{p})\tau_3+\hat{\Delta}\tau_1e^{-i\phi\tau_3}\label{eq:SynthHam}
\end{align}

\noi that is similarly defined in momentum-phase $(\bm{p},\phi)$ space. 

Hence, under the assumption of a weak and uniform $\bm{\nabla}\phi$, and by employing the above newly defined quantities, we find that the current per volume which flows in the $i$-th direction due to a phase gradient imposed in the $j$-th direction, takes the compact form:
\begin{align}
J_i=-\partial_j\phi\int dP\ph{\rm Tr}\Big[\hat{\upsilon}_i(\bm{p})\mathds{1}_\tau\hat{\cal F}_{p_j\phi}(\epsilon,\bm{p},\phi)\Big].
\label{eq:StartingPoint}
\end{align}

\noi Equations~\eqref{eq:SynthGreen}-\eqref{eq:StartingPoint} are the key relations for the reformulation of superfluid response at zero temperature and key general results of this work. Extensions to finite tem\-pe\-ra\-tu\-re are straightforward by considering the finite temperature Matsubara Green function framework~\cite{AltlandSimons}.

\section{Adiabatic Approach:\\ Application to 1D Systems}\label{sec:Adiab1D}

In the remainder, we apply the above formalism to va\-rious superconducting systems. We begin by con\-si\-de\-ring strictly 1D SCs and demonstrate how the superfluid transport can be viewed as a manifestation of 1D chiral anomaly. Subsequently, we proceed with 2D systems and demonstrate how the arising quantization of the superfluid stiffness can be understood through either dimensional extension of the 1D chiral anomaly, or, the emergence of genuinely 2D topological effects.

\subsection{1D Superconducting Dirac Cone}

Our first case study concerns a superconducting Dirac cone in 1D as described in Sec.~\ref{sec:TIedge} and is experimentally realizable on the edge of a 2D spin Hall insulator with proximity induced conventional superconductivity. Within our adia\-ba\-tic framework, the resulting synthetic space ``single-particle'' Hamiltonian obtains the form:
\begin{align}
\hat{{\cal H}}(p_x,\phi)=\upsilon_Dp_x\tau_3+\Delta\tau_1e^{-i\phi\tau_3}\,.
\label{eq:AdiabaticHam1D}
\end{align}

\noi The above Hamiltonian features an antiunitary charge-conjugation symmetry which is effected by the operator $\hat{\Xi}=\tau_3{\cal K}$, where $\hat{{\cal K}}$ defines the operation of complex conjugation in synthetic space, i.e., it inverts both momentum and phase. Consequently, the synthetic space Hamiltonian belongs to symmetry class D, and can be in principle characterized by a $\mathbb{Z}$ topological invariant~\cite{Schnyder2008}, which is associated with the 1st Chern number $C_1$ of the occupied band~\cite{Niu}. This is given by:
\begin{align}
C_1=\int dp_x\int_0^{2\pi}\frac{d\phi}{2\pi}\ph\Omega_{p_x\phi}(p_x,\phi)\label{eq:firstChern}
\end{align}

\noi where we introduced the Berry curvature $\Omega_{p_x\phi}(p_x,\phi)$ of the occupied band. We remark that, in general, the 1st Chern number is quantized according to $C_1\in\mathbb{Z}$. This happens under the condition that $p_x$ is defined in a compact space, which is obviously not the case here since $p_x\in(-p_c,+p_c)$. However, for the case of an odd number of Dirac electron branches and the cutoff momentum $p_c$ taken to infinity, one still obtains that $C_1\in\mathbb{Z}$~\cite{CA}. This is a manifestation of chiral anomaly and stems from the fact that the phase $\phi$ which twists the Dirac mass does not enter the energy spectrum. In this case, $|C_1|$ counts the number of Dirac points in the band structure.

\subsubsection{Reformulated Theory of Superfluid Transport - Topological Pumping}\label{sec:TopoPump1D}

The emergence of chiral anomaly and the topological nature of superfluid transport {\color{black}is described} more naturally using the reformulated theory introduced in this work. Since for the present model $\hat{\upsilon}_x(p_x)=\upsilon_D$, Eq.~\eqref{eq:StartingPoint} implies that the current for $B=0$ takes the transparent form:
\begin{align}
J_x=-\upsilon_D\int_{-p_c}^{+p_c}\frac{dp_x}{2\pi}\ph\Omega_{p_x\phi}(p_x,\phi)\,\partial_x\phi\,,\label{eq:JxTI}
\end{align}

\noi where we introduced the Berry curvature:
\begin{align}
\Omega_{p_x\phi}(p_x,\phi)=\int_{-\infty}^{+\infty}\frac{d\epsilon}{2\pi}\ph{\rm Tr}\Big[\hat{\cal F}_{p_x\phi}(\epsilon,p_x,\phi)\Big],
\end{align}

\noi which is defined in the synthetic $(p_x,\phi)$ space. Since we assume that $B=0$, the Berry curvature takes contributions only from the occupied band of the Hamiltonian in Eq.~\eqref{eq:AdiabaticHam1D}. Relations similar to Eq.~\eqref{eq:JxTI} are typical for 1D chiral anomaly~\cite{Kotetes} and to\-po\-lo\-gi\-cal pumps~\cite{Thouless}. The connection to the former is established by noticing that $\int_{-p_c}^{+p_c}dp_x\ph\Omega_{p_x\phi}(p_x,\phi)=1$ for $p_c\rightarrow\infty$, thus allowing us to reach once again to the quantization of the superfluid stiffness in units of the Fermi velocity, i.e., $D_{xx}=\upsilon_D/\pi$.

We remind the reader that the current $J_x$ defines the current per length of the 1D SC. Therefore, we can further define the current $I_x$ which flows through a finite-sized system with length $L_x$, across which, $\phi(x)$ becomes modified by $\Delta\phi$. Hence, by integrating Eq.~\eqref{eq:JxTI} over the $x$ coordinate for a finite length of the system $L_x$, we find that the total current flowing in the $x$ direction is:
\begin{align}
I_x=\int_{-L_x/2}^{+L_x/2}dx\,J_x=-\upsilon_D\int_{\phi(-L_x/2)}^{\phi(+L_x/2)}\frac{d\phi}{2\pi}=-\upsilon_D\frac{\Delta\phi}{2\pi}\,.
\end{align}

\noi Thus, when $\Delta\phi$ is ($\pi$) $2\pi$, $I_x$ becomes (fractionally) quantized in units of $\upsilon_D$. Interestingly, Eq.~\eqref{eq:JxTI} defines a Thouless pump in coordinate space~\cite{Niu}, in analogy to the usual Thouless pump defined in the time domain~\cite{Thouless}.

\subsubsection{Emergence of Chiral Anomaly}\label{sec:ChiralAnomaly}

The quantization of the superfluid stiffness can be alternatively attributed to the emergence of chiral anomaly. For 1D Dirac electrons, the realization of chiral anomaly is manifested in the particular form of the effective action $S$ which describes the respective U(1) scalar $a_0$ and vector $a_1$ potentials, along with their chiral analogs, the U(1) scalar $b_0$ and vector $b_1$ chiral gauge potentials. Specifically, by integrating out the massless/massive Dirac electrons  defined in $d=1$, the effective action one obtains for the two types of U(1) gauge fields reads as~\cite{Semenoff,CA,Kotetes}:
\begin{align}
S=-\frac{1}{\pi}\iint dtdx\,\varepsilon_{\mu\nu}a^\mu b^\nu\,.
\end{align}

\noi The above action is expressed using the relati\-vi\-stic coor\-di\-na\-te vector $x^\mu=(t,x)$ and the metric tensor $\eta_{\mu\nu}={\rm diag}\{1,-1\}$, where $\mu,\nu=0,1$. We also introduced the antisymmetric Levi-Civita symbol $\varepsilon_{\mu\nu}$, while summation of repeated Greek indices is implied throughout.

The phenomenon of chiral anomaly dictates the nonconservation of the chiral charge even when the Dirac electrons become massless. In particular, the chiral two-current is defined as $j_b^\mu=-\delta S/\delta b_\mu$, and is given by the expression:
\begin{align}
j_b^{\mu}=-\frac{1}{\pi}\varepsilon^{\mu\nu}a_\nu\,.
\end{align}

\noi The fact that chiral charge is not conserved is reflected in the relation:
\begin{align}
\partial_\mu j_b^{\mu}=-\frac{1}{\pi}\varepsilon^{\mu\nu}\partial_\mu a_\nu\equiv-\frac{{\cal E}_x}{\pi}\,.
\end{align}

\noi In the above, we made use of the fact that in $d=1$ the term $\varepsilon^{\mu\nu}\partial_\mu a_\nu$ is equivalent to the electric field ${\cal E}_x$ of the respective U(1) gauge field. On the other hand, the U(1) two-current is given by the analogous expression:
\begin{align}
j_a^{\mu}=\frac{1}{\pi}\varepsilon^{\mu\nu}b_\nu\,.
\end{align}

\noi In spite of the obvious similarity arising for the expressions obtained for the usual and chiral currents, the U(1) charge is conserved, since the chiral gauge fields $b_\mu$ can be generally expressed in the form $b_\mu=\partial_\mu\varphi/2$. Here, $\varphi$ constitutes the phase which twists the Dirac mass. This specific property that is sa\-ti\-sfied by the chiral gauge fields is vital for obtaining the local conservation law of the U(1) charge, i.e., $\partial_\mu j_a^{\mu}=0$.

The discussion and results regarding 1D chiral anomaly directly apply to the present situation. To make the connection clear, it is first required to identify the usual and chiral U(1) gauge fields. First of all, we remark that the phase $\varphi$ coincides here with the superconducting phase $\phi$. In addition, we find that the U(1) chiral gauge fields are given by the expressions:
\begin{align}
b^0=V-\partial_t\phi/2\quad{\rm and}\quad
b^1=A_x+\partial_x\phi/2\,.
\end{align}

\noi Therefore, due to the oppositely charged electrons and holes, the usual electromagnetic potentials play here the role of U(1) chiral gauge fields. This was already pointed out earlier, e.g., in Ref.~\onlinecite{StoneLopes}. As a result, the U(1) gauge fields are here given by the expressions:
\begin{align}
a^0=-\upsilon_D\big(A_x+\partial_x\phi/2\big)\quad{\rm and}\quad a^1=-\big(V-\partial_t\phi/2\big)/\upsilon_D\,.
\end{align}

\noi The antisymmetric relation satisfied by the usual and chiral U(1) gauge fields is crucial to recover the action of a 1D SC which reads as:
\begin{align}
S=\frac{\upsilon_D}{2\pi}\iint dtdx\Big[\big(V-\partial_t\phi/2\big)^2/\upsilon_D^2-\big(A_x+\partial_x\phi/2\big)^2\Big].\label{eq:1DChiralAnomalyAction}
\end{align}

\noi The above is manifestly gauge invariant, as required for a SC, and gives rise to the electrostatic and Meis\-sner screening effects~\cite{AltlandSimons}. Even more, differen\-tia\-ting the action with respect to $A_x$ also allows us to recover the result we obtained earlier in Eq.~\eqref{eq:Current1D} for the current $J_x$.

Besides recovering the quantization of the superfluid stiffness, the underlying role of chiral anomaly implies that there exists an additional physical phenomenon which accompanies the quantization of superfluid stiffness. To identify the associated physical quantity which also becomes quantized in ``suitable units'' we consider the derivative with respect to $V$. The latter yields the excess charge density:
\begin{align}
\rho_c=-\frac{\partial S}{\partial V}=\frac{1}{\upsilon_D}\frac{\partial_t\phi}{2\pi}\,.
\end{align}

\noi Since a time-dependence in the phase can be induced in a Josephson junction by a scalar potential $V$, $\rho_c$ can be in principle detected as the excess charge density de\-ve\-lo\-ping across the voltage-biased Josephson junction. As we discuss in detail in Sec.~\ref{sec:JQC}, the JQC which is defined after $c_{\cal Q}=\rho_c/(\partial_t\phi/2)$, constitutes the chiral anomaly partner quantity of the superfluid stif\-fness. As such, it is also expected to exhibit topological phenomena.

\begin{figure}[t!]
\begin{center}
\includegraphics[width=\columnwidth]{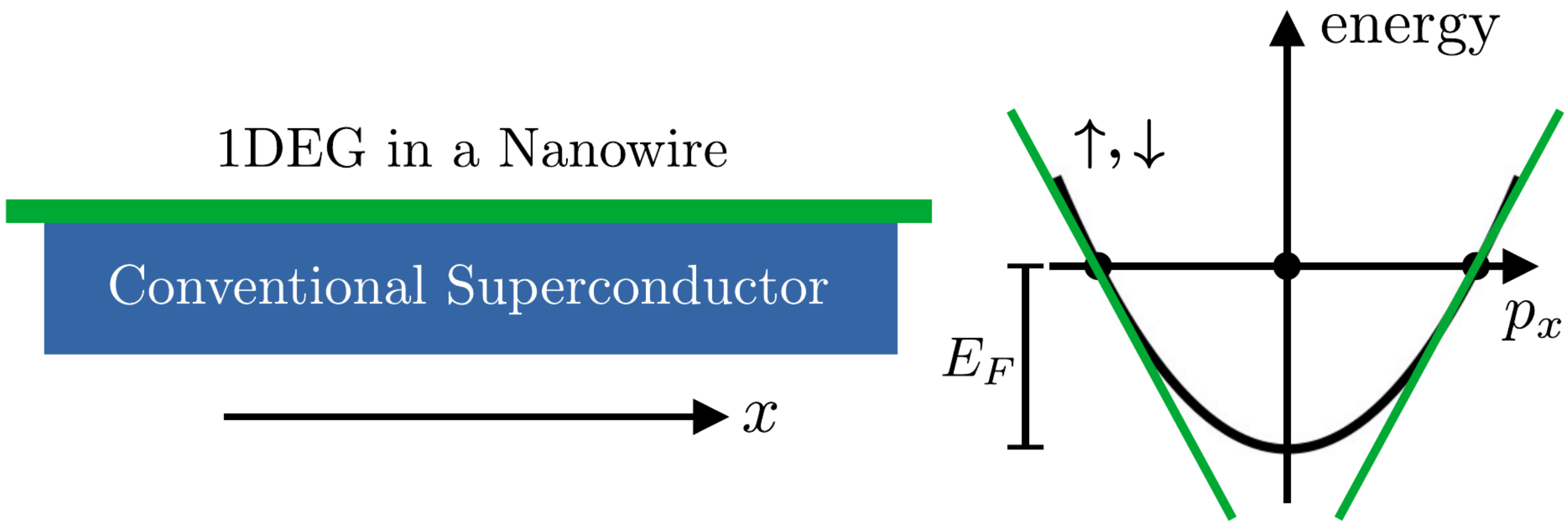}
\end{center}
\caption{Conventional superconductor interfacing a one-dimensional electron gas (1DEG) confined in a single-channel quantum nanowire. When the Fermi level of the 1DEG is sufficiently larger than the proximity-induced pairing gap on the nanowire, the quadratic energy dispersion of the nanowire can be linearized about the two Fermi points $\pm p_F$. Hence, within the linear dispersion approximation, the superfluid stiffness of the superconducting nanowire is twice the stiffness of the superconducting helical edge modes in Fig.~\ref{fig:PRRFigure1}. This is because the 1DEG carries both spin degrees of freedom $\uparrow,\downarrow$.}
\label{fig:PRRFigure2}
\end{figure}

\subsection{Superconducting 1D Electron Gas}

The above results are not restricted to pristine Dirac systems, such as topologically-protected boun\-dary modes, but are also applicable to platforms which exhibit an emergent Dirac behavior. For example, this is the case for a 1D electron gas which is described by the quadratic energy dispersion $\hat{h}(p_x)=p_x^2/2m-E_F$, where $E_F$ defines the Fermi energy in the normal phase.

Indeed, a 1D electron gas can effectively demonstrate Dirac physics in the so-called quasiclassical limit where $E_F\gg\Delta$ holds. In this limit, the dispersion can be linea\-ri\-zed {\color{black}about each Fermi point $\pm p_F$ lying at} ener\-gy $E_F$. See also Fig.~\ref{fig:PRRFigure2} for an illustration. This results in right and left mover electrons with dispersions {\color{black}($\upsilon_F=p_F/m$)}:
\begin{align}
\hat{h}_\pm(p_x)=\pm\upsilon_F(p_x\mp p_F)\,.
\end{align}

\noi Based on our calculation for the supercurrent carried by the topological edge of a spin Hall insulator with pro\-xi\-mi\-ty induced conventional pairing, we here obtain that:
\begin{align}
J_x^{\rm 1DEG}=-2\upsilon_F\frac{\partial_x\phi}{2\pi}\,.
\end{align}

\noi Notably, the above result coincides with the current obtained for a 1D Josephson junction in the long junction limit~\cite{Furusaki}.

\section{Adiabatic Approach:\\ Application to 2D Systems}\label{sec:Adiab2D}

After exemplifying our approach for 1D supercon\-duc\-ting semimetals, we now employ Eq.~\eqref{eq:StartingPoint} to obtain the superfluid stif\-fness for a 2D superconducting Dirac cone. As it is also pointed out in our work in Ref.~\onlinecite{WJA}, this result also allows us to explain the quantized outcome for the superfluid stiffness found in Ref.~\onlinecite{KopninSoninPRL} for 2D superconduc\-ting monolayer graphene in the Dirac-cone regime. As we show in the upcoming paragraphs, the quantized contribution of the Dirac-cone part of the band structure to the superfluid stiffness can be either understood by exten\-ding to 2D the conclusions obtained from the emergence of chiral anomaly in 1D, or, by directly accounting for the topological properties of the 2D Dirac Hamiltonian.

\subsection{Superfluid Stiffness of a Superconducing 2D Dirac Cone as a Result of 1D Chiral Anomaly}

We now consider the description of superfluid transport using our new approach and evaluate the superfluid stif\-fness of a single 2D Dirac cone by means of the expression in Eq.~\eqref{eq:StartingPoint}. Follo\-wing this route allows us to naturally expose the underlying role of the phenomenon of 1D chiral anomaly discussed earlier. To proceed, we first introduce the respective adiabatic Hamiltonian for a single superconducting Dirac cone in 2D with $\hat{\Delta}=\Delta\mathds{1}_\sigma$:
\begin{align}
\hat{\cal H}(\bm{p},\phi)=\upsilon_D(p_x\sigma_1+p_y\sigma_2)\tau_3+\Delta\tau_1e^{-i\phi\tau_3},\label{eq:2DDiracCone}
\end{align}

\noi and re-express it in the limit of small $\phi$ according to:
\begin{align}
\hat{\cal H}(\bm{p},\phi)=\hat{\cal U}(p_y)\big[\upsilon_Dp_x\sigma_1\tau_3+m(p_y)\tau_1-\Delta\phi\tau_2\big]\hat{\cal U}^\dag(p_y),\no
\end{align}

\noi where we introduced the effective Dirac mass $m(p_y)=\sqrt{(\upsilon_D p_y)^2+\Delta^2}$, and the unitary matrix:
\[\hat{{\cal U}}(p_y)={\rm Exp}\big[i\gamma(p_y)\sigma_2\tau_2/2\big]\]

\noi with a phase $\gamma(p_y)$ which is given by the defining relation $\cos[\gamma(p_y)]=\Delta/m(p_y)$. Using the above, we now transfer to a new frame with:
\bea
\hat{\cal H}'(\bm{p},\phi)&=&\hat{\cal U}^\dag(p_y)\hat{{\cal H}}(\bm{p},\phi)\hat{\cal U}(p_y)\no\\
&=&\upsilon_Dp_x\sigma_1\tau_3+m(p_y)\tau_1-\Delta\phi\tau_2\,.
\eea

\noi The Hamiltonian is block diagonal in the new frame, since it commutes with $\sigma_1$. The same property holds for $\hat{\cal F}'_{p_x\phi}(\epsilon,\bm{p},\phi)$ and the respective matrix Berry curvature $\hat{\Omega}_{p_x\phi}'(\bm{p},\phi)=\Omega_{p_x\phi}'(\bm{p},\phi)\sigma_1$. Note that the quantity $\Omega_{p_x\phi}'(\bm{p},\phi)$ corresponds to the Berry curvature of the ne\-ga\-ti\-ve energy band of the $\sigma_1=1$ block of $\hat{\cal H}'(\bm{q},\phi)$.

To proceed, we assume that $\phi\ll1$ and find the expression~\cite{VolovikBook,Niu}:
\begin{align}
\Omega_{p_x\phi}'(\bm{p},\phi)=\frac{1}{\cos[\gamma(p_y)]}\frac{\upsilon_D\Delta^2}{2\sqrt{(\upsilon_Dp_x)^2+\big[m(p_y)\big]^2}^3}.\label{eq:BerryCurvBand}
\end{align}

\noi The matrix structure of $\hat{\Omega}_{p_x\phi}'(\bm{p},\phi)\propto\sigma_1$ reflects that it belongs to the Euler class~\cite{Ahn,Senthil1,Senthil2}. Consequently, tra\-cing it over the $\sigma$ sublattice indices yields zero. Therefore, convo\-lu\-ting the Berry curvature with the normal phase Bloch electron group velocity operator in the new frame defined as:
\bea
\hat{\upsilon}_x'(\bm{p})&=&\hat{\cal U}^\dag(p_y)\upsilon_D\sigma_1\hat{\cal U}(p_y)\no\\
&=&\upsilon_D\cos[\gamma(p_y)]\sigma_1-\upsilon_D\sin[\gamma(p_y)]\sigma_3\tau_2\,,\no
\eea

\noi is crucial to obtain a nonzero current, in analogy to the nonlinear Hall effect induced by Berry dipoles~\cite{Sodemann,Xie,Ortix}.

Indeed, also here we have dipoles consisting of Berry monopoles with charges $\sigma_1=\pm1$. These are Weyl points in synthetic $(p_x,\phi,m(p_y))$ space with locations identified by the singularities of $\Omega_{p_x\phi}'(\bm{p},\phi)$. In Eq.~\eqref{eq:BerryCurvBand}, the Weyl point locations are independent of $\phi$, as it is customary for topological responses governed by chiral anomaly~\cite{Kotetes}.

Since $\hat{\cal F}'_{p_x\phi}(\epsilon,\bm{p},\phi)\propto\sigma_1$, only the part of $\hat{\upsilon}_x'(\bm{p})$ which is proportional to $\sigma_1$ contributes. Therefore, by assuming $B=0$, Eq.~\eqref{eq:JxTI} implies that the current becomes:
\begin{align}
\frac{J_x}{\partial_x\phi}=-\int_{-p_c}^{+p_c}\frac{dp_y}{\pi}\ph\upsilon_D\cos[\gamma(p_y)]\int_{-p_c}^{+p_c}\frac{dp_x}{2\pi}\ph\Omega_{p_x\phi}'(\bm{p},\phi)\,.
\label{eq:JxGraphene}
\end{align}

\noi The evaluation of Eq.~\eqref{eq:JxGraphene} for $p_c\rightarrow\infty$ provides in a straightforward fashion:
\begin{align}
J_x=-\frac{\Delta}{2\pi}\partial_x\phi\,,\label{eq:Quantization2D}
\end{align}

\noi as a result of 1D chiral anomaly occuring for an infinite set of uncoupled sectors each of which is labeled by the transverse momentum $p_y$.

\subsection{Quantization due to Nontrivial Topology in 2D}

The result of the previous paragraph is certainly remar\-ka\-ble, since the quantization effects encountered in the purely 1D chiral anomaly also persist when con\-si\-de\-ring a 2D Dirac system. Such a result cannot be a coincidence, but should be instead well-rooted to the properties of the synthetic Dirac Hamiltonian in Eq.~\eqref{eq:2DDiracCone} which dictates the superfluid transport of a superconducting Dirac cone in 2D.

To reveal the underlying reason for this quantization by means of a genuinely 2D point of view, let us first analyze in further depth the topological properties of the respective 2D Hamiltonian. We find that the Hamiltonian in Eq.~\eqref{eq:2DDiracCone} is identical to the one that was first discussed by Jackiw and Rossi~\cite{JackiwRossi}. Specifically, it possesses a chiral symmetry $\{\hat{\cal H}(\bm{p},\phi),\hat{\Pi}\}=\hat{0}$, which is generated by the action of the operator $\hat{\Pi}=\sigma_3\tau_3$. Moreover, one addi\-tio\-nally finds antiunitary symmetries~\cite{Altland1997,Schnyder2008}. Specifically, the Hamiltonian is invariant under the action of a charge conjugation and a generalized TR transformation, which are effected by the operators $\hat{\Xi}=\sigma_1\tau_3\hat{{\cal K}}$ and $\hat{{\cal T}}=i\sigma_2\hat{{\cal K}}$, respectively. We remind the reader that invariance under charge conjugation implies that $\big\{\hat{{\cal H}}(\bm{p},\phi),\hat{\Xi}\big\}=\hat{0}$, while invariance under TR results in $\big[\hat{{\cal H}}(\bm{p},\phi),\hat{{\cal T}}\big]=\hat{0}$.

From this symmetry analysis, we conclude that the synthetic Hamiltonian belongs to class DIII. Hence, it can be topologically classified using a win\-ding number denoted $w_3$~\cite{Schnyder2008}. The latter is an integer when the synthetic space is compactified. Remarkably, however, for $\phi\in[0,2\pi)$ and the Dirac Hamiltonian in question, the win\-ding number also takes integer values, i.e., $w_3\in\mathbb{Z}$. Note that this is in spite of the fact that the base space is not compact. We remind the reader that a similar be\-ha\-vior was observed for the 1st Chern number in Eq.~\eqref{eq:firstChern}, and can be also here attributed to the fact that the ener\-gy spectrum is independent of $\phi$.

This topological invariant can be expressed in terms of the Hamiltonian $\hat{\cal H}(\bm{p},\phi)$ in the follow fashion:
\begin{align}
w_3=\frac{\varepsilon_{ijk}}{48\pi^2}\int d^3\tilde{p}\ph{\rm Tr}\Big(\hat{\Pi}\ph\hat{\cal H}^{-1}\partial_{\tilde{p}_i}\hat{\cal H}\ph\hat{\cal H}^{-1}\partial_{\tilde{p}_j}\hat{\cal H}\ph\hat{\cal H}^{-1}\partial_{\tilde{p}_k}\hat{\cal H}\Big),\label{eq:winding3}
\end{align}

\noi where we introduced the synthetic momentum vector $(\tilde{p}_1,\tilde{p}_2,\tilde{p}_3)=(p_x,p_y,\phi)$ and the antisymmetric Levi-Civita symbol $\varepsilon_{ijk}$, where $i,j,k=1,2,3$. The phase integration over $\phi$ takes place in the interval $\phi\in[0,2\pi)$, while one is expected to extend the integration over $\bm{p}$ in all real numbers in the plane $\mathbb{R}^2$. The above winding number predicts the appearance of zero modes pinned by vortices induced in the Dirac mass field $\Delta(\bm{r})e^{i\phi(\bm{r})}$, as it was first proposed by Jackiw and Rossi~\cite{JackiwRossi}. Moreover, it also dictates the emergence of Majorana zero modes in the celebrated Fu-Kane model~\cite{FuKane} when superconducting vortices are introduced on the surface of a 3D to\-po\-lo\-gi\-cal insulator. Inte\-re\-stin\-gly, the same invariant predicts the pinning of Majorana zero modes by vortices introduced in magnetic texture crystals, which interface nodal SCs~\cite{SteffensenVortex}.

At this point, it is important to stress that when the Hamiltonian is of the Dirac type, the phase which involves the twisting of the mass field in coordinate space does not enter the ener\-gy spectrum and, as a result, it does not appear in the brackets of Eq.~\eqref{eq:winding3} after evalua\-ting the various derivatives. Hence, quantization effects do not only arise for $w_3$ but also emerge for the winding number density defined according to expression:
\begin{align}
w_3(\phi)=\int \frac{d\bm{p}}{2\pi}\,w_3(\bm{p},\phi),
\label{eq:winding3density}
\end{align}

\noi where we introduced the winding number density in the full 3D synthetic space:
\begin{align}
w_3(\bm{p},\phi)=\frac{1}{2}\frac{\varepsilon_{ijk}}{3!}{\rm Tr}\Big(\hat{\Pi}\ph\hat{\cal H}^{-1}\partial_{\tilde{p}_i}\hat{\cal H}\ph\hat{\cal H}^{-1}\partial_{\tilde{p}_j}\hat{\cal H}\ph\hat{\cal H}^{-1}\partial_{\tilde{p}_k}\hat{\cal H}\Big).
\label{eq:winding3densitymomentum}
\end{align}

\noi As a matter of fact, an analogous relation holds for $J_x$ and its integrated counterpart $I_x$, as shown in Sec.~\ref{sec:TopoPump1D}, with the former mapping to $w_3(\phi)$ and the latter to $w_3$. In the case of a superconducting Dirac cone in 2D, we find that:
\begin{align}
w_3(\bm{p},\phi)=-\frac{2\upsilon_D^2\Delta^2}{E^4(\bm{p})}\,.\label{eq:winding3densityDirac}
\end{align}

\noi Plugging the above in Eq.~\eqref{eq:winding3density}, leads to $w_3(\phi)=-1$.

We now proceed with the main goal of this section, which is to demonstrate that the quantization found in Eq.~\eqref{eq:Quantization2D} emerges due to the fact that the dia\-go\-nal element of the superfluid stiffness tensor for supercon\-duc\-ting graphene in the Dirac regime is related to $w_3$. To prove this, we start from Eq.~\eqref{eq:StartingPoint} and make use of the relations $D_{xx,yy}=D\Rightarrow D=\big(D_{xx}+D_{yy}\big)/2$, to write:
\begin{align}
D=\sum_{i=x,y}\int dP\ph{\rm Tr}\Big[\hat{\upsilon}_i(\bm{p})\mathds{1}_\tau\hat{\cal F}_{p_i\phi}(\epsilon,\bm{p},\phi)\Big]\,.\label{eq:StiffnessDirac2Dw3}
\end{align}

\noi We now make use of the relation $\partial_\epsilon\hat{{\cal G}}^{-1}=i$ along with $\hat{\upsilon}_{x,y}(\bm{p})=\upsilon_D\sigma_{1,2}$, in order to carry out the substitutions:
\begin{align}
\hat{\upsilon}_x(\bm{p})\partial_\epsilon\hat{{\cal G}}^{-1}=-\hat{\Pi}\hat{\upsilon}_y(\bm{p})\tau_3\equiv+\hat{\Pi}\partial_{p_y}\hat{{\cal G}}^{-1},\\
\hat{\upsilon}_y(\bm{p})\partial_\epsilon\hat{{\cal G}}^{-1}=+\hat{\Pi}\hat{\upsilon}_x(\bm{p})\tau_3\equiv-\hat{\Pi}\partial_{p_x}\hat{{\cal G}}^{-1}.
\end{align}

\noi By plugging the above result into Eq.~\eqref{eq:StiffnessDirac2Dw3} we find~\cite{WJA}:
\bea
D&=&\frac{1}{2}\int dP\,{\rm Tr}\Big[\hat{\Pi}\,\big(\partial_{p_x}\hat{\cal G}^{-1}\big)\hat{\cal G}\,\big(\partial_{p_y}\hat{\cal G}^{-1}\big)\hat{\cal G}\,\big(\partial_\phi\hat{\cal G}^{-1}\big)\hat{\cal G}\Big]\no\\
&-&\frac{1}{2}\int dP\,{\rm Tr}\Big[\hat{\Pi}\,\big(\partial_{p_x}\hat{\cal G}^{-1}\big)\hat{\cal G}\,\big(\partial_{\phi}\hat{\cal G}^{-1}\big)\hat{\cal G}\,\big(\partial_{p_y}\hat{\cal G}^{-1}\big)\hat{\cal G}\Big]\no\\
&+&\frac{1}{2}\int dP\,{\rm Tr}\Big[\hat{\Pi}\,\big(\partial_{p_y}\hat{\cal G}^{-1}\big)\hat{\cal G}\,\big(\partial_{\phi}\hat{\cal G}^{-1}\big)\hat{\cal G}\,\big(\partial_{p_x}\hat{\cal G}^{-1}\big)\hat{\cal G}\Big]\no\\
&-&\frac{1}{2}\int dP\,{\rm Tr}\Big[\hat{\Pi}\,\big(\partial_{p_y}\hat{\cal G}^{-1}\big)\hat{\cal G}\,\big(\partial_{p_x}\hat{\cal G}^{-1}\big)\hat{\cal G}\,\big(\partial_{\phi}\hat{\cal G}^{-1}\big)\hat{\cal G}\Big].\no\\
\label{eq:StiffnessGraphene2DG}
\eea

\noi We observe that the above is missing two more sequences of derivatives in order to complete all six possible permutations of the form \[\varepsilon_{ijk}\big(\partial_{\tilde{p}_i}\hat{\cal G}^{-1}\big)\hat{\cal G}\,\big(\partial_{\tilde{p}_j}\hat{\cal G}^{-1}\big)\hat{\cal G}\,\big(\partial_{\tilde{p}_k}\hat{\cal G}^{-1}\big)\hat{\cal G}\,.\]

\noi However, it is straightforward to confirm that the re\-mai\-ning two terms can be obtained from the existing terms. Indeed, the sequences $(\partial_{p_x},\partial_{p_y},\partial_\phi)$ and $(\partial_{p_y},\partial_{\phi},\partial_{p_x})$ are equivalent to the sequence $(\partial_\phi,\partial_{p_x},\partial_{p_y})$, while the sequences $(\partial_{p_x},\partial_\phi,\partial_{p_y})$ and $(\partial_{p_y},\partial_{p_x},\partial_{\phi})$ are equivalent to $(\partial_\phi,\partial_{p_y},\partial_{p_x})$. Hence, by suitably converting parts of the existing terms into the missing ones, we can write:
\begin{align}
D=\frac{\varepsilon_{ijk}}{3}\int dP\,{\rm Tr}\Big[\hat{\Pi}\,\big(\partial_{\tilde{p}_i}\hat{\cal G}^{-1}\big)\hat{\cal G}\,\big(\partial_{\tilde{p}_j}\hat{\cal G}^{-1}\big)\hat{\cal G}\,\big(\partial_{\tilde{p}_k}\hat{\cal G}^{-1}\big)\hat{\cal G}\Big].
\no\\\label{eq:SSperm}
\end{align}

At this stage we can further simplify the above expression and reveal its connection to $w_3$. For this purpose, we take into account that $\big[\hat{\cal H}(\bm{p},\phi)\big]^2=E^2(\bm{p})\mathds{1}$, i.e., all positive/negative energies are given by $\pm E(\bm{p})$ where $E^2(\bm{p})=\sqrt{\varepsilon^2(\bm{p})+\Delta^2}$ with $\hat{h}^2(\bm{p})=\varepsilon^2(\bm{p})\mathds{1}_h$. See also Sec.~\ref{sec:SSTopoMetals}. Given the above, we carry out the integral over energy $\int_{-\infty}^{+\infty}d\epsilon/2\pi$, and after {\color{black}the manipulations discussed in  Appendix~\ref{app:AppendixD} we} obtain:
\begin{align}
D=\int\frac{d\bm{p}}{(2\pi)^2}\,w_3(\bm{p},\phi)\,E^3(\bm{p})\frac{d}{dE(\bm{p})}\left\{\frac{\Theta\big[E(\bm{p})-|B|\big]}{E(\bm{p})}\right\}
\label{eq:SSfinal}
\end{align}

\noi which can be further expanded to yield:
\bea
D&=&-\int\frac{d\bm{p}}{(2\pi)^2}\,w_3(\bm{p},\phi)\,E(\bm{p})\,\Theta\big[E(\bm{p})-|B|\big]\no\\
&&+\int\frac{d\bm{p}}{(2\pi)^2}\,w_3(\bm{p},\phi)\,B^2\,\delta\big[E(\bm{p})-|B|\big]\,.
\label{eq:SSfinalexpanded}
\eea

\noi From the above we observe that the superfluid stiffness for $B=0$ simplifies to:
\begin{align}
D_{B=0}=-\int\frac{d\bm{p}}{(2\pi)^2}\,w_3(\bm{p},\phi)\,E(\bm{p})\,.
\label{eq:SSfinalexpandedZeroB}
\end{align}

It is straightforward to confirm that also the above expression leads to a quantized superfluid stiffness which is equal to $\Delta/\pi$. The above expression clearly demonstrate that the value of the superfluid stiffness is set by the winding number density which, in turn, is also fixed by the topological properties of the STS. {\color{black}Even more, as we show in Appendix~\ref{app:AppendixE}, the above formula can be ge\-ne\-ra\-li\-zed to provide the result of Eq.~\eqref{eq:HOTBTP}.}

The above shown topological nature of the superfluid stiffness further implies its stability to perturbations which preserve chiral symmetry. In our companion work in Ref.~\onlinecite{WJA}, we also examine the resilience of the superfluid stiffness against chiral-symmetry preserving types of perturbations. Among these, we also investigate the inclusion of disorder in the modulus $\Delta$, which renders it spatially dependent, i.e., $\Delta(\bm{r})$. Our analysis shows that the superfluid stiffness retains the same form, but with $\Delta$ now being replaced by a spatially averaged pairing gap.

\section{Topological Aspects of Josephson Quantum Capacitance}\label{sec:JQC}

As it was found when exploring the superfluid stiffness of 1D systems, see for instance Sec.~\ref{sec:Adiab1D}, the realization of 1D chiral anomaly also implies that the quantum ca\-pa\-ci\-tance constitutes another quantity which is expected to exhibit quantization phe\-no\-me\-na. Since we are dea\-ling with superconducting systems, we are here inte\-re\-sted in the study of the quantum capacitance arising in Josephson junctions, or JQC as we refer to it. We are essentially interested in junctions whose two superconducting leads are separated by a highly efficient dielectric, so that the system acts as a capacitor. In this limit, the Josephson coupling becomes negligible and charge builds up on each superconducting plate.

\begin{figure}[t!]
\begin{center}
\includegraphics[width=0.94\columnwidth]{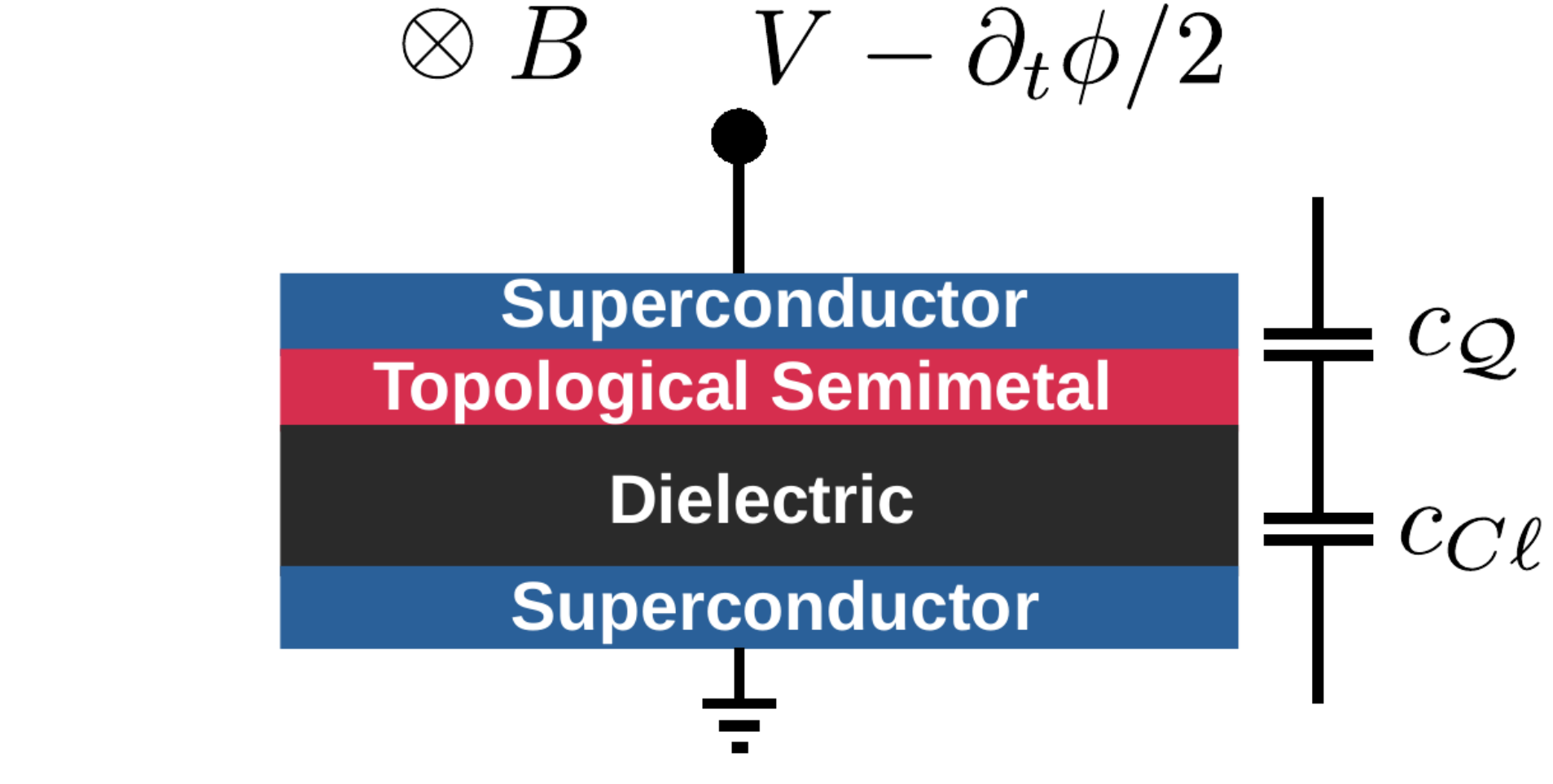}
\end{center}
\caption{Hybrid system for the measurement of the Josephson quantum capacitance (JQC). The dielectric is assumed to of high efficiency so that the Josephson coupling across the junction is fully suppressed and the heterostructure functions as a capacitor. There exist two contributions to the junction capacitance: the classical ($c_{C\ell}$) and the quantum capacitance ($c_{\cal Q}$). The latter stems from degrees of freedom of the to\-po\-lo\-gical semimetal which dictate the low energy sector of the heterostructure and experience a proximity induced pairing gap. This is under the assumption of a sufficiently weak voltage bias or rate for phase time variations, which is much smaller than the supercoducting gap of the bulk supercoductors. Lastly, the classical contribution to the capacitance depends on the characteristics of the junction, and originates from the charge response of the large number of electrons which are occupied below the Fermi level in the metallic leads.}
\label{fig:PRRFigure3}
\end{figure}

In the cases of inte\-rest, we consider that two superconducting leads sandwich the topological semimetal and the dielectric, thus, leading to a lateral SC-topological semimetal-dielectric-SC heterostructure as shown in Fig.~\ref{fig:PRRFigure3}. In such a system, the low-energy degrees of freedom stem from the topological semimetal which sees a proximity induced gap. We consider that the superconducting gap of the superconducting semimetal, which is controlled by one of the two conventional supercoducting leads, picks up a time-dependent phase $\phi(t)$.

It is eligible to restrict to the phase of the STS, since we can assume that the phase of the other supercon\-duc\-ting lead, which is attached to the dielectric, is set to zero. Hence, $\phi$ corresponds to the phase difference appearing across the junction, that is further assumed to be biased by a voltage bias $V$. Gauge invariance implies that the electrostatic potential and the superconducting phase appear together accor\-ding to $V\mapsto V-\partial_t\phi/2$. The above coupling naturally leads to the generation of excess charge density $\rho_c$ for a nonzero $\partial_t\phi$. Note that $\rho_c$ does not include the charge density stemming from the electronic states of the two superconducting metallic leads which lie energetically sufficiently below the Fermi level $E_F$, which is here assumed to satisfy $E_F\gg |V-\partial_t\phi/2|$.

For a constant $\partial_t\phi$, there is an additional contribution to the capacitive energy per area of the Josephson junction $E_{\rm JJ}$, which reads as $E_{\rm JJ}=-c_{\rm JJ}\big(V-\partial_t\phi/2\big)^2/2$, cf Eq.~\eqref{eq:1DChiralAnomalyAction}. Here, $c_{\rm JJ}$ denotes the total capacitance per area of the Josephson junction, which includes the classical ($c_{C\ell}$) and quantum ($c_{\cal Q}$) parts. Since the two capacitances are in series, we have the relation:
\begin{align}
\frac{1}{c_{\rm JJ}}=\frac{1}{c_{C\ell}}+\frac{1}{c_{\cal Q}}\,.
\end{align}

\noi The classical capacitance is controlled by the geometric properties of the junction and is in principle tunable by modifying the design parameters of the heterostructure. Therefore, by rendering $c_{C\ell}$ much larger than $c_{\cal Q}$, we can essentially eliminate the influence of the former.

In the following, we first review the standard approach to theo\-re\-tical\-ly evaluating the quantum capacitance and, afterwards, we provide a reformulation which tran\-spa\-ren\-tly exposes the emergence of topological effects.

\subsection{Standard Theory}

The JQC is inferred by evaluating the charge susceptibility of the junction, i.e., $c_{\cal Q}=-\partial^2 E_{\rm JJ}/\partial V^2$. By restricting to the case $\big[\hat{\Delta},\hat{h}(\bm{p})\big]=\hat{0}$, linear response yields:
\bea
&&c_{\cal Q}=-\int dP\ph{\rm Tr}\Big[\tau_3\hat{G}(\epsilon,\bm{p})\tau_3\hat{G}(\epsilon,\bm{p})\Big]\no\\
&&=\sum_\alpha\int\frac{d\bm{p}}{(2\pi)^d}\,\left\{1-\left[\frac{\Delta_\alpha(\bm{p})}{B}\right]^2\right\}\delta\big[E_\alpha(\bm{p})-|B|\big]\no\\
&&+\sum_\alpha\int\frac{d\bm{p}}{(2\pi)^d}\,\frac{\Delta_\alpha^2(\bm{p})}{E_\alpha^3(\bm{p})}\,P_\alpha(\bm{p}),
\label{eq:JQCformulaDif}
\eea

\noi where $\alpha$ labels the eigenstates of $\hat{h}(\bm{p})$ with dispersions $\varepsilon_\alpha(\bm{p})$ and pairing gap $\Delta_\alpha(\bm{p})=\big<u_\alpha(\bm{p})\big|\hat{\Delta}\big|u_\alpha(\bm{p})\big>$. Hence, we end up with the Bogoliubov quasiparticle ener\-gy $E_\alpha(\bm{p})=\sqrt{\varepsilon_\alpha^2(\bm{p})+\Delta_\alpha^2(\bm{p})}$. Moreover, we employed the band defined parity $P_\alpha(\bm{p})=\Theta\big[E_\alpha(\bm{p})-|B|\big]$, which has already been discussed in Appendix~\ref{app:AppendixB}. Note that for $|\Delta_\alpha(\bm{p})|>|B|$ the parity of the respective band becomes equal to unity. Thus, the terms in the se\-cond row of Eq.~\eqref{eq:JQCformulaDif} are nonzero only for $|B|>|\Delta_\alpha(\bm{p})|$, i.e., when the Bogoliubov-Fermi level set by the Zeeman ener\-gy scale $B$ crosses the bands and the system is metallic. It is convenient to rewrite the above using the normal phase density of states. For this purpose, we define the energy $E(h)=\sqrt{h^2+\Delta^2(h)}$, and write:
\bea
c_{\cal Q}&=&\int_{-\infty}^{+\infty}dh\, \varrho(h)\,\bigg\{\frac{\Delta^2(h)}{E^3(h)}\,P(h)\no\\
&&\qquad\quad+\Big\{1-\big[\Delta(h)/B\big]^2\Big\}\delta\big[E(h)-|B|\big]\bigg\}\,\,\quad\label{eq:JQC}
\eea

\noi where we introduced the normal-phase density of states:
\begin{align}
\varrho(h)=\sum_\alpha\int\frac{d\bm{p}}{(2\pi)^d}\,\delta\big[\varepsilon_\alpha(\bm{p})-h\big]\,.
\end{align}

\noi In addition, we introduced the parity $P(h)=\Theta\big[E(h)-|B|\big]$ which is obtained after the replacement $E_\alpha(\bm{p})\mapsto E(h)$. Finally, we note that in the event that $\Delta(h)=\Delta$, the expression for the JQC can be compactly expressed in the following fashion:
\begin{align}
c_{\cal Q}=\int_{-\infty}^{+\infty}dh\, \varrho(h)\,\frac{d}{dh}\left[\frac{hP(h)}{E(h)}\right]\,.
\end{align}

\subsection{Adiabatic Reformulation}

In analogy to our adiabatic approach employed for the superfluid stiffness, here we need to consider temporal variations of the superconducting phase. For this purpose, we consider the time-dependent Hamiltonian:
\begin{align}
\hat{H}(t,h)=h\tau_3+\Delta\tau_1e^{-i\phi(t)\tau_3},\label{eq:HamTime}
\end{align}

\noi which is expressed in terms of $h$ and an $h$-independent fixed value for the pairing gap $\Delta$. To obtain the excess charge density $\rho_c$, we eva\-lua\-te the expectation value of the electric charge operator $\hat{\rho}_c=-\tau_3$ in response to $\partial_t\phi$. In analogy to the steps considered for the case of superfluid stiffness in Sec.~\ref{sec:StiffnessReform}, we also here start from the Dyson equation and take into account the first order correction to the single-particle matrix Green function $\hat{G}(\epsilon,h)$ due to the perturbation term $-\Delta\phi(t)\tau_2$. The bare Green function is here defined according to the relation $\hat{G}^{-1}(\epsilon,h)=i\epsilon+B-\hat{H}(h)$ with $\hat{H}(h)=h\tau_3+\Delta\tau_1$.

To proceed, we consider a Wick rotation $\tau=it$ to imaginary time, we transfer to Fourier space $\tau\mapsto\omega$ where $\omega$ denotes the imaginary energy, and we end up with the symmetrized expression for the ensuing Dyson equation:
\begin{align}
\hat{G}^{(1)}(\epsilon,\omega,h)\approx\frac{\hat{G}(\epsilon,h)}{2}\left[2\pi\delta(\omega)-\Delta\phi(\omega)\tau_2\hat{G}(\epsilon-\omega,h)\right]\no\\
\quad\qquad\qquad+\left[2\pi\delta(\omega)-\hat{G}(\epsilon+\omega,h)\Delta\phi(\omega)\tau_2\right]\frac{\hat{G}(\epsilon,h)}{2}.
\end{align}

\noi Since here we are in\-te\-re\-sted in an adiabatic approach, we restrict to the insulating regime, and assume slow temporal variations for $\phi(t)$ so that $|\omega|\ll\Delta$. These assumptions allow us to take the limit $|\omega|\rightarrow0$. By further con\-si\-de\-ring a uniform rate $\partial_t\phi$, we obtain the approximate time-independent modified matrix Green function:
\bea
\hat{G}^{(1)}(\epsilon,h)&\approx&\hat{G}(\epsilon,h)+\partial_t\phi\,\hat{G}(\epsilon,h)\frac{\Delta\tau_2}{2}\partial_\epsilon\hat{G}(\epsilon,h)\no\\
&&-\partial_t\phi\,\big[\partial_\epsilon\hat{G}(\epsilon,h)\big]\frac{\Delta\tau_2}{2}\hat{G}(\epsilon,h)\,.
\eea

Having identified the perturbed Green function, we now move ahead and obtain the expectation value for the excess charge density:
\begin{align}
\rho_c=-\int_{-\infty}^{+\infty}dh\,\varrho(h)\int_{-\infty}^{+\infty}\frac{d\epsilon}{2\pi}\ph{\rm Tr}\left[\delta\hat{G}^{(1)}(\epsilon,h)\tau_3\right]_{\Delta\mapsto\Delta(h)},
\end{align}

\noi where we substracted the background charge density $\sim{\rm Tr}\big[\hat{G}(\epsilon,h)\tau_3\big]$. Therefore, in the above we employ the first order correction of the matrix Green function $\delta\hat{G}^{(1)}(\epsilon,h)=\hat{G}^{(1)}(\epsilon,h)-\hat{G}(\epsilon,h)$ in terms of $\partial_t\phi$. By replacing the correction with its explicit form, we find:
\bea
&&\rho_c=\partial_t\phi\int_{-\infty}^{+\infty}dh\,\varrho(h)\int_{-\infty}^{+\infty}\frac{d\epsilon}{2\pi}\ph{\rm Tr}\Big\{\no\\
&&\big[\partial_\epsilon\hat{G}(\epsilon,h)\big]\Big[\Delta\tau_2\hat{G}(\epsilon,h)\tau_3-\tau_3\hat{G}(\epsilon,h)\Delta\tau_2\Big]/2\Big\}_{\Delta\mapsto\Delta(h)}.\no
\eea

\noi Within the limit of a weak $\partial_t\phi$ examined here, we write the last row in the above expression in the following form:
\bea
&&\big[\partial_\epsilon\hat{G}(\epsilon,h)\big]\Big[\Delta\tau_2\hat{G}(\epsilon,h)\tau_3-\tau_3\hat{G}(\epsilon,h)\Delta\tau_2\Big]/2\no\\
&&\qquad\approx e^{-i\phi\tau_3/2}\hat{\cal G}(\epsilon,h,\phi)\hat{{\cal F}}_{h\phi}(\epsilon,h,\phi)\hat{\cal G}^{-1}(\epsilon,h,\phi)e^{i\phi\tau_3/2}\quad\no
\eea

\noi where we introduced the matrix function $\hat{\cal F}_{h\phi}(\epsilon,h,\phi)$:
\begin{align}
\hat{\cal F}_{h\phi}=\nicefrac{1}{2}\big(\partial_\epsilon\hat{\cal G}^{-1}\big)\hat{\cal G}\big(\partial_\phi\hat{\cal G}^{-1}\big)\hat{\cal G}\big(\partial_h\hat{\cal G}^{-1}\big)\hat{\cal G}-\partial_\phi\leftrightarrow\partial_{h}\,,
\end{align}

\noi along with the matrix Green function:
\begin{align}
\hat{\cal G}^{-1}(\epsilon,h,\phi)=i\epsilon+B-\hat{\cal H}(h,\phi)\,,\label{eq:SynthGreenQtime}
\end{align}

\noi which is defined in the synthetic $(\epsilon,h,\phi)$ space and results from the Hamiltonian:
\begin{align}
\hat{\cal H}(h,\phi)=h\tau_3+\Delta\tau_1e^{-i\phi\tau_3}\,.
\label{eq:SynthHamQtime}
\end{align}

By means of the above manipulations we end up with the following expression for the JQC:
\begin{align}
c_{\cal Q}\equiv\frac{\rho_c}{\partial_t\phi/2}=2\int_{-\infty}^{+\infty}dh\, \varrho(h)\,\Omega_{h\phi}(h,\phi)\,,
\label{eq:ReformQ}
\end{align}

\noi where we introduced the synthetic space Berry curvature of the occupied bands:
\begin{align}
\Omega_{h\phi}(h,\phi)=\int_{-\infty}^{+\infty}\frac{d\epsilon}{2\pi}\,{\rm Tr}\Big[\hat{\cal F}_{h\phi}(\epsilon,h,\phi)\Big]_{\Delta\mapsto\Delta(h)}.\label{eq:SyntheticBerryQtime}
\end{align}

Notably, we find that in analogy to Eqs.~\eqref{eq:SynthGreen}-\eqref{eq:StartingPoint}, here Eqs.~\eqref{eq:SynthGreenQtime}-\eqref{eq:SyntheticBerryQtime} are key for the adiabatic reformulation of the zero-temperature JQC. Once again, extensions to finite temperature are straightforward. Finally, we also remark that eva\-lua\-ting the above synthetic Berry curvature is straightforward, and allows us to recover the standard expression in Eq.~\eqref{eq:JQC} when the insulating regime is strictly considered.

\subsection{Applications}

In the following paragraphs we demonstrate how the above applies to the variety of STSs exa\-mi\-ned earlier. In all the following cases we exa\-mi\-ne the scenario of $\Delta(h)=\Delta$ and $B=0$, in which event one obtains $P(h)=1$.

\subsubsection{1D Superconducting Topological Semimetals}

To evaluate the JQC, we need to first obtain the density of states. Here, there is only one band with energy dispersion $\varepsilon(p_x)=\upsilon_Dp_x$ and we have:
\begin{align}
 \varrho(h)=\int_{-\infty}^{+\infty}\frac{dp_x}{2\pi}\ph\delta(\upsilon_Dp_x-h)=\frac{1}{2\pi\upsilon_D}\,.
\end{align}

It is straightforward to obtain the JQC, thanks to the emergence of chiral anomaly, in which case the Berry curvature $\Omega_{h\phi}(h,\phi)$ is independent of $\phi$, and its integral counts the number of touching points. Therefore, we find:
\begin{align}
c_{\cal Q}=\frac{1}{\pi\upsilon_D}\,.
\end{align}

\noi Interestingly, for 1D STSs, the product of the superfluid stiffness and the JQC obtains a universal value, that is:
\begin{align}
Dc_{\cal Q}=\frac{1}{\pi^2},
\end{align}

\noi where we set for simplicity $D=D_{xx}$. The above result can be viewed as a constitutive relation for 1D STSs.

\subsubsection{2D Superconducting Topological Semimetals}

We now proceed with investigating the JQC for a single 2D Dirac cone. Here, one finds two bands with energy dispersions $\varepsilon_\pm(p)=\pm\upsilon_Dp$ and we have:
\bea
&& \varrho(h)=
\int\frac{d\bm{p}}{(2\pi)^2}\ph\sum_{\sigma=\pm1}\delta(\sigma\upsilon_Dp-h)\no\\
&&=
\int_0^{\infty}\frac{dp\, p}{2\pi}\ph\big[\delta(\upsilon_Dp-h)+\delta(\upsilon_Dp+h)\big]=\frac{|h|}{2\pi\upsilon_D^2}\,.\no\\
\eea

\noi By employing Eq.~\eqref{eq:ReformQ}, we find that the JQC in the present case reads as:
\bea
c_{\cal Q}
&=&\frac{1}{\pi\upsilon_D^2}\int_{-\infty}^{+\infty}dh\,|h|\,\Omega_{h,\phi}(h,\phi)\no\\
&=&\frac{1}{\pi\upsilon_D^2}\int_0^\infty dh\, h\,\frac{\Delta^2}{\sqrt{h^2+\Delta^2}^3}=\frac{\Delta}{\pi\upsilon_D^2}\,.
\eea

\noi Also for this class of systems we are in a position to obtain a constitutive relation linking superfluid stiffness and the JQC, which reads as:
\begin{align}
Dc_{\cal Q}=\left(\frac{\Delta}{\pi\upsilon_D}\right)^2\,,
\end{align}

\noi and involves the superconducting coherence length $\xi_{\rm sc}$ of the STS which is given by $\xi_{\rm sc}=\upsilon_D/\Delta$.

\section{Influence of the Zeeman Field}\label{sec:Zeeman}

Up to this point, our analysis focused on the case $B=0$, in which only the negative energy bands are occupied. Since the Zeeman energy $B$ plays the role of a chemical potential and sets the Bogoliubov-Fermi level, it is important to investigate its influence on the superfluid stiffness and the JQC for the two representative systems studied earlier.

\subsubsection{1D Superconducting Topological Semimetals}

By employing Eq.~\eqref{eq:StiffnessMain}, we find that for an arbitrary value of $B$, the superfluid stiffness for the system in exa\-mi\-ned in Sec.~\ref{sec:TIedge} takes the form:

\begin{align}
D_{xx}=\frac{\upsilon_D}{\pi}\left[1-\frac{\Theta\big(|B|-\Delta\big)}{\sqrt{1-\big(\Delta/B\big)^2}}\right].
\end{align}

\noi The above implies that the superfluid stiffness is discontinuous across $|B|=\Delta$. i.e., when the Dirac band tou\-ching point is crossed. Notably, in the limit $|B|\rightarrow\infty$, we find that $D_{xx}$ goes to zero.

We now proceed with examining the impact of modifying the energy level occupancy on the JQC. Using the expression in Eq.~\eqref{eq:JQC}, we find that $c_{\cal Q}=1/\pi\upsilon_D$, that is, it is independent of the Zeeman field. While the robustness of the JQC against arbitrary Zeeman field variations is remarkable, it also implies that it is impossible to observe any distinctive features of the underlying Dirac BTP by means of controlling this external control knob. Hence, it is only the investigation of the superfluid stiffness across the $|B|=\Delta$ point that can yield characteristic signatures of the STS.

\subsubsection{2D Superconducting Topological Semimetals}

Repeating the same procedure for the case of the single Dirac cone in two spatial dimensions of Sec.~\ref{eq:2dDiracCone}, leads to the diagonal superfluid stiffness $D=D_{xx,yy}$:\footnote{In fact, we can alternatively obtain the expression for $D$ using Eq.~\eqref{eq:SSfinal} in conjunction with Eq.~\eqref{eq:winding3densityDirac}.}
\begin{align}
D=\frac{\Delta}{\pi}\Theta\big(\Delta-|B|\big)\,.
\end{align}

\noi Notably, as soon as the Zeeman energy exceeds the pai\-ring gap, the superfluid stiffness vanishes. This remar\-ka\-ble result highlights that the entire superfluid stif\-fness is carried by the Dirac BTP,  which in the supercon\-duc\-ting phase is split at energies $\pm\Delta$. Therefore, the superfluid stiffness yields a smoking gun signature of the STS upon varying the Zeeman energy. In fact, this property is crucial for disentangling the pre\-sen\-ce of the Dirac BTP in a band structure which is not described by the ideal Dirac cone model. Our partner work in Ref.~\onlinecite{WJA} discusses how this fingerprint can be employed to infer the Dirac BTPs from the superfluid stiffness of superconducting graphene.

The respective JQC can be obtained by employing Eq.~\eqref{eq:JQC} and leads to the expression:
\begin{align}
c_{\cal Q}=\frac{{\rm max}\big\{\Delta,|B|\big\}}{\pi\upsilon_D^2}\,.
\end{align}

\noi Notably, the variation of the JQC with respect to $|B|$ can also reflect the presence of the underlying BTP in 2D. Indeed, while $c_{\cal Q}$ remains continuous across $|B|=\Delta$ its derivative $dc_{\cal Q}/d|B|$ exhibits a jump of $1/\pi\upsilon_D^2$.

{\color{black}

\subsubsection{Experimental Feasibility of the Desired Zeeman Control}

}
Concluding this {\color{black}section, it is important to stress once more} that signatures of STSs are obtained for Zeeman energies larger than the pairing gap. {\color{black} However, a Zeeman field is known to have a dramatic effect on spin-singlet superconductivity, since it leads to net magnetization which tends to break Cooper pairs. Hence, it is not ob\-vious that the desired condition $|B|=\Delta$ can be met in experiments.}

\begin{figure}[t!]
\begin{center}
\includegraphics[width=0.87\columnwidth]{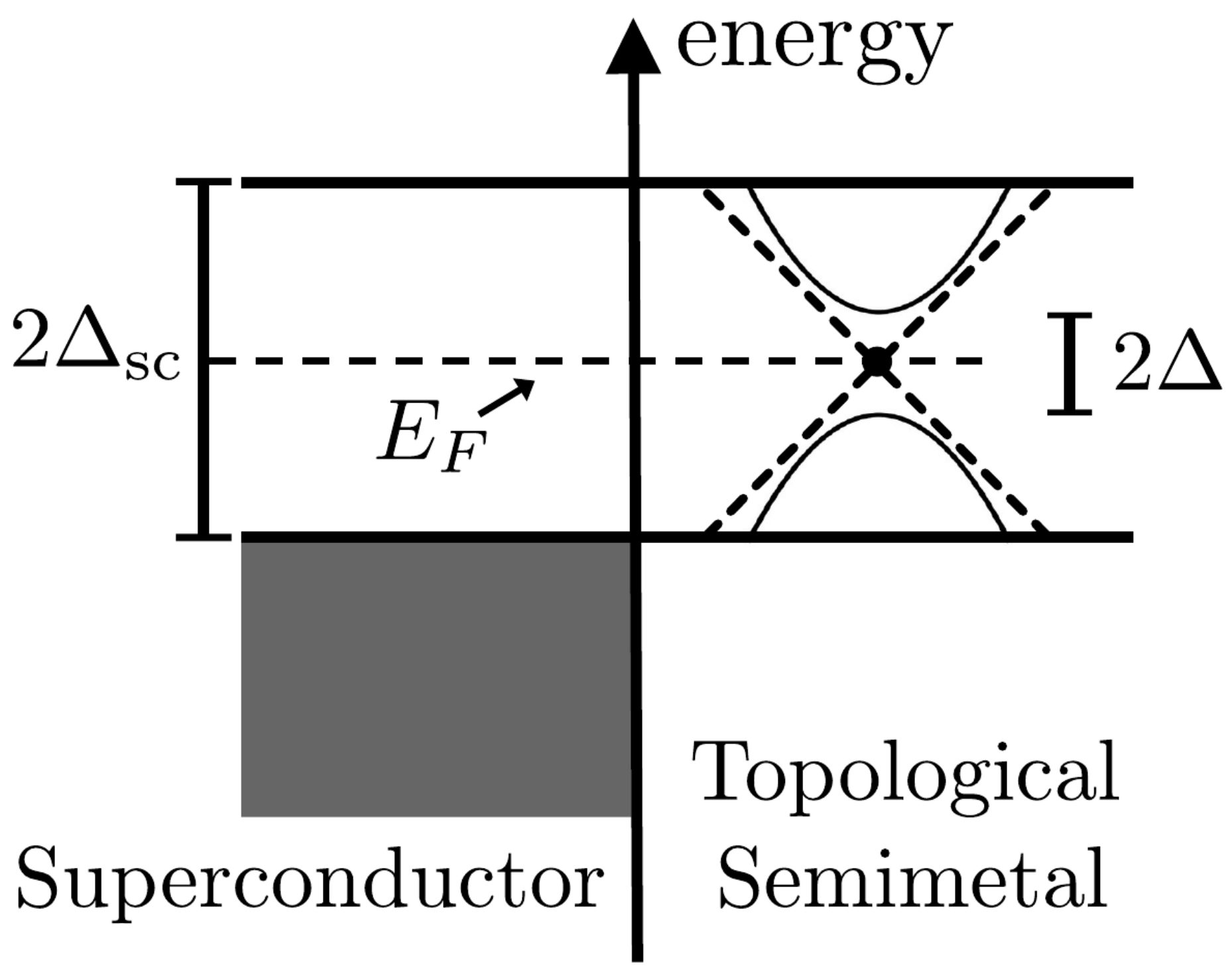}
\end{center}
\caption{Band diagram which describes the proximity effect between a conventional metallic superconductor and a topological semimetal. As follows from Refs.~\onlinecite{Antipov,Mikkelsen}, the Fermi level of the superconductor sets the Fermi level of the entire hybrid system. The proximity induced gap on the semimetal $\Delta$ is typically smaller than the pairing gap $\Delta_{\rm sc}$ in the bulk superconductor. Hence, there is a window for which one can control the contribution of the band touching point of the topological semimetal to the superfluid stiffness and quantum capacitance using a Zeeman field, without modifying the respective contributions originating from the electrons of the bulk superconductor.}
\label{fig:PRRFigure4}
\end{figure}

For intrinsic SCs, such a regime is challen\-ging to achieve experimentally, since uniform superconducti\-vi\-ty cannot be su\-stained for high Zeeman fields. {\color{black}Spe\-ci\-fi\-cal\-ly, for a thin film intrinsic STS, superconductivity is expected to be destroyed when the magnetic energy scale reaches the Chandrasekhar-Clogston limit (CCL)~\cite{CCL}, i.e., $B_{\rm CC}=\Delta/\sqrt{2}$. An alternative possi\-bi\-li\-ty is that the system de\-ve\-lops a} spatially-modulated so-called Fulde-Ferrell-Larkin-Ovchinikov supercon\-duc\-ting ground state~\cite{FF,LO} before reaching the $B_{\rm CC}$ value. {\color{black}In either case, the here-predicted phenomena appear to be expe\-ri\-men\-tal\-ly inacessible or at least very difficult to achieve in intrinsic supercoductors.}

However, such an obstacle can be circumvented for topological se\-mi\-me\-tals which expe\-rien\-ce a pairing gap {\color{black}$\Delta$} inherited by means of proximity from a bulk SC, which plays the role of a Cooper pair bath. The proximity induced pairing on the topological semimetal is {\color{black}ge\-ne\-ral\-ly smaller} than the bulk superconducting gap {\color{black}$\Delta_{\rm sc}$. For instance, following the analysis of the superconducting proximity effect in Ref.~\cite{Potter}, we can write the induced pairing gap on the topological semimetal in terms of the pairing gap of the parent superconductor according to $\Delta=\big(1-{\cal Z}\big)\Delta_{\rm sc}\leq\Delta_{\rm sc}$. Here, ${\cal Z}\in[0,1]$ denotes the renorma\-li\-za\-tion factor resulting from the coupling between the supercoductor and the semimetal.

To observe the discontinuities in the superfluid stiffness and JQC discussed in the previous sections, the condition $|B|=\Delta$ needs to be satisfied. At the same time, $|B|$ has to be smaller than the CCL of the parent superconductor, i.e., $|B|<\Delta_{\rm sc}/\sqrt{2}$, so that superconductivity is sustained in the entire hybrid system. The above con\-si\-de\-ra\-tions imply that the minimum value ${\cal Z}_{\rm min}$ of ${\cal Z}$ that is required for $|B|=\Delta$ to be met is ${\cal Z}_{\rm min}\simeq0.3$. Hence, this relatively low required value for ${\cal Z}_{\rm min}$ implies that for high-quality interfaces with a ${\cal Z}$ much larger than ${\cal Z}_{\rm min}$,} there should be a window for which the Zeeman energy can exceed the pairing gap in the STS while remaining safely below the {\color{black}CCL for the parent superconductor.

Lastly, we conclude by providing in Fig.~\ref{fig:PRRFigure4} a band dia\-gram which describes} the proximity effect between the conventional superconductor and {\color{black}a topological semimetal}. For further details on the role of the band alignment on the proximity effect, see Refs.~\onlinecite{Antipov,Mikkelsen}.\\

\section{Conclusions and Outlook}\label{sec:Conclusions}

In this work we unify the diagonal superfluid responses of spin-singlet superconductors which are characterized by a fully gapped bulk energy spectrum. We find that for superconductors with emergent Lorentz in\-va\-riance, the superfluid stiffness $D$ and quantum capacitance $c_{\cal Q}$ satisfy the relation $D=\upsilon_D^2c_{\cal Q}$~\cite{WJA}, where $\upsilon_D$ defines the ensuing ``speed of light". The above naturally arises in Dirac-type superconductors, which in their normal phase contain topological band touching points and crossings.

Even more importantly, we show that such superconducting topological semimetals further exhibit to\-po\-lo\-gical effects which stem from the nontrivial topological charge of these special points in the normal phase band structure. Both $D$ and $c_{\cal Q}$ become proportional to a topological invariant quantity, which counts the number of such special points in the band structure. Hence, the two quantities become ``quantized" but in units which depend on the material parameters, such as, the speed of light and the pairing gap. Nonetheless, in spite of the nonuniversal character of these quantized effects, their to\-po\-lo\-gi\-cal nature renders them robust against weak uncorrelated disorders~\cite{WJA}. Moreover, the expressions found for these coefficients in the clean case, also carry over in the disor\-dered case, with the only difference that now the bulk material parameters need to be replaced by their disorder-averaged counterparts~\cite{WJA}.

Main goal of this work is to reveal the underlying to\-po\-lo\-gi\-cal nature of these response coefficients, and introduce a suitable general framework to study and identify such topological diagonal responses. For this purpose, we show that viewing the superfluid stiffness and quantum capacitance as the charge current and density responses induced by spatial and temporal va\-ria\-tions of the superconducting phase, allows expressing them in terms of Berry curvatures defined in appropriate synthetic spaces. This approach provides a natural explanation for the resulting topological quantization, since it attributes it to the topological charge of the singularities of these synthetic Berry curvatures. Even more, we show that for one- and two-dimensional superconducting topological semimetals, the emergence of the quantization can be understood as the outcome of chiral anomaly.

The present and our accompanying work in Ref.~\onlinecite{WJA} set the stage for the exploration of topological diagonal superfluid responses and bring the measurements of the superfluid stiffness and the Josephson quantum ca\-pa\-ci\-tan\-ce as a means of diagnosing the presence of Berry singularities in the system's band structure. Therefore, these two quantities can be viewed as a particular type of Berry singularity makers. The concept of Berry singularity markers was earlier introduced in Refs.~\onlinecite{BSM1,BSM2}. This method relies on extracting information regar\-ding the presence of topological band touching points in a band structure by investigating the diagonal responses of the system. Nonetheless, our work does not only promise to motivate further theoretical developments, but it can also guide experimentalists to observe the here-found to\-po\-lo\-gi\-cal effects.

In fact, the observation of the quantization effects discussed in the main text {\color{black}are in principle experimentally} feasible in superconductor-graphene hybrids~\cite{Morpurgo,Andrei,Choi,Mizuno,Calado,Finkelstein,Shalom,Bretheau,Seredinski}, in which graphene inherits a conventional supercon\-duc\-ting gap due to the pro\-xi\-mi\-ty effect. {\color{black}Currently, however, it is very challenging to tune the chemical potential $\mu$ of graphene sufficiently close to the Dirac point so that $|\mu|\ll\Delta$~\cite{Mayorov}. Therefore, future fabrication and technological advancements are required for achie\-ving this goal. Nonetheless, in our companion paper~\cite{WJA} we discuss that although the stiffness is not topologically quantized when $\mu$ is switched on, the superfluid stif\-fness in the antipodal and experimentally accessible limit $|\mu|\gg\Delta$ remains proportional to the absolute value of the vorticity of the band touching point, see also Appendix~\ref{app:AppendixE}. As a result, this observation opens a prominent route to study part of the topological aspects brought forward in this work with presently accessible experimental platforms.}

{\color{black}At this point, it is crucial to} remark that in rea\-li\-stic Dirac-type materials and hybrids, there exist additional contributions to the two coefficients besides the ones studied here. These originate from the non\-re\-la\-ti\-vi\-stic regions of the band structure, and they tend to spoil the here-found quantized effects. In order for the experimentalists to be in a position to disentangle the desired contribution of the topological band touchings and cros\-sings to the two quantities of interest, we propose to externally apply a magnetic field, which couples only through the Zeeman effect to the system. In this event, the sole but yet crucial function of the magnetic field, is to set the Bogoliubov-Fermi level of the system. In fact, in the main text, we show that the superfluid stiffness and the Zeeman-field-derivative of the Josephson quantum ca\-pa\-ci\-tan\-ce ge\-ne\-ral\-ly exhibit discontinuities when the Zeeman energy scale exceeds the pairing gap. {\color{black} Meeting this condition in experiments, however, appears to be a challenging task. For intrinsic superconducting topological semimetal, superconductivity is expected to be either already destroyed in lower fields~\cite{CCL}, or, converted into an unconventional Fulde-Ferrell-Larkin-Ovchinikov phase~\cite{FF,LO}. On the other hand, hybrid platforms appear more suitable for testing these phenomena, since we find that such a condition is possible to satisfy, even for interfaces with a moderate proximity effect.

Before concluding this work, it is imperative to emphasize that our predictions for the topological superfluid stiffness and the Josephson quantum capacitance are solely applicable to one- and two-dimensional superconducting  topological semimetals. As we have already briefly mentioned in Sec.~\ref{sec:3DSTS}, these superfluid responses are not capable of capturing the topological charge of Weyl points appearing in three-dimensional topological semimetals. The reason why these superfluid stiffness cannot be employed to reflect the topological charges of band touching points  in higher dimensions, is due to the same reason for which the ``strong" topological properties of a three-dimensional system cannot be captured by a lower-dimensional topological invariant. In ge\-ne\-ral, topological systems can be classified into hier\-ar\-chies which are related by dimensional extension and reduction~\cite{Ryu,Qi}. For instance, Chern insulators in two and four dimensions are classified by the first and second Chern numbers $C_1$ and $C_2$, and belong to two distinct hierarchies. This implies that the topological properties of insulators obtained by dimensional reduction and/or extension in each hierarchy can be linked to $C_1$ and $C_2$.

In the present case, the two topological superfluid responses in one and two spatial dimensions become linked because, in a similar sense, these superconducting topological semimetals belong to the same hierarchy. In one (two) dimension(s) we find that the superfluid stiffness is related to the synthetic-space $C_1$ Chern ($w_3$ winding) number. As we also show, the two-dimensional case can be also understood by extending chiral anomaly to two dimensions. The three-dimensional case, however, belongs to a different hierarchy for which we expect that a synthetic $C_2$ Chern (or a $w_5$ winding) number is relevant. Hence, the here-discussed current-current responses are not capable of exposing a $C_2$ or a $w_5$ charge, since the latter can be only associated with a higher-order current correlation function. Nonetheless, the metho\-do\-lo\-gy and adiabatic formalism introduced in this work lay the foundations for the study of higher-order response functions that could potentially classify superconducting topological semimetals in higher dimensions. Hence, our approach sets the stage for the further exploration and discovery of phenomena in time-reversal superconductors which originate from nontrivial synthetic $(\bm{p},\phi)$ topology.}

\section*{Acknowledgements}

We are thankful to Mario Cuoco, Maria Teresa Mercaldo, Tao Shi, and Hong-Qi Xu for helpful discussions. We acknowledge funding from the National Na\-tu\-ral Scien\-ce Foundation of China (Grant No.~12074392).

\appendix

\section{Details on the Derivation of the Standard Superfluid Stiffness Formula}\label{app:AppendixA}

\noi We now re-express the paramagnetic current contribution (first row) of the result shown in Eq.~\eqref{eq:Dprefinal}, and obtain the following formula for the paramagnetic contribution $D_{ij}^{(p)}$ to the superfluid stiffness:
\bea
D_{ij}^{(p)}&=&\int dP\ph{\rm Tr}\Big[\hat{\upsilon}_j(\bm{p})\tau_3\tau_3\hat{G}(\epsilon,\bm{p})\tau_3\hat{\upsilon}_i(\bm{p})\tau_3\hat{G}(\epsilon,\bm{p})\Big]\no\\
&=&\int dP\ph{\rm Tr}\Big\{\hat{\upsilon}_j(\bm{p})\tau_3\big[\tau_3,\hat{G}(\epsilon,\bm{p})\big]\hat{\upsilon}_j(\bm{p})\mathds{1}_\tau\hat{G}(\epsilon,\bm{p})\no\\
&&\quad\ph+\partial_{p_j}\hat{G}^{-1}(\epsilon,\bm{p})\hat{G}(\epsilon,\bm{p})\partial_{p_i}\hat{G}^{-1}(\epsilon,\bm{p})\hat{G}(\epsilon,\bm{p})\Big\}.\no
\eea

It is straightforward to show that the second contribution to the paramagnetic term $D_{ij}^{(p)}$ is opposite to the diamagnetic one, thus cancelling each other out. This can be made transparent by rewriting the diamagnetic contribution as follows:
\bea
&&D_{ij}^{(d)}=\int dP\ph{\rm Tr}\Big[\hat{G}(\epsilon,\bm{p})\partial_{p_jp_i}\hat{H}(\bm{p})\Big]\no\\
&&\qquad\,\equiv
\int dP\ph{\rm Tr}\Big[\partial_{p_j}\hat{G}(\epsilon,\bm{p})\partial_{p_i}\hat{G}^{-1}(\epsilon,\bm{p})\Big]\no\\
&&=-\int dP\ph{\rm Tr}\Big[\partial_{p_j}\hat{G}^{-1}(\epsilon,\bm{p})\hat{G}(\epsilon,\bm{p})\partial_{p_i}\hat{G}^{-1}(\epsilon,\bm{p})\hat{G}(\epsilon,\bm{p})\Big],\no
\eea

\noi where we employed the relation $\partial\hat{G}=-\hat{G}\partial\hat{G}^{-1}\hat{G}$. To obtain the second line we used a partial integration and relied on the fact that momentum is defined in a compact space, e.g., a BZ.

\section{Band-Defined Superfluid Stiffness}\label{app:AppendixB}

To arrive to an equivalent representation which assigns a superfluid stiffness contribution to each band, one introduces the eigenstates of $\hat{h}(\bm{p})$, which we label as $\big|u_\alpha(\bm{p})\big>$ with energy dispersions $\varepsilon_\alpha(\bm{p})$. Under the assumption $\big[\hat{h}(\bm{p}),\hat{\Delta}\big]=\hat{0}$ we also set $E_\alpha(\bm{p})=\sqrt{\varepsilon_\alpha^2(\bm{p})+\Delta_\alpha^2(\bm{p})}$ with the band defined pairing gap $\Delta_\alpha(\bm{p})=\big<u_\alpha(\bm{p})\big|\hat{\Delta}\big|u_\alpha(\bm{p})\big>$. After carrying out the integration of Eq.~\eqref{eq:StiffnessMain} over energy, we end up with the band-index dependent formula:
\bea
D_{ij}&=&2\int_{\rm BZ}\frac{d\bm{p}}{(2\pi)^d}\sum_{\alpha,\beta}\Delta_\alpha(\bm{p})\Delta_\beta(\bm{p})\left[\frac{P_\beta(\bm{p})}{E_\beta(\bm{p})}-\frac{P_\alpha(\bm{p})}{E_\alpha(\bm{p})}\right]\no\\
&&\cdot\frac{\big<u_\alpha(\bm{p})\big|\partial_{p_i}\hat{h}(\bm{p})\big|u_\beta(\bm{p})\big>\big<u_\beta(\bm{p})\big|\partial_{p_j}\hat{h}(\bm{p})\big|u_\alpha(\bm{p})\big>}{E_\alpha^2(\bm{p})-E_\beta^2(\bm{p})}\,,\no\\
\eea

\noi where we introduced the parity function for a given Bogoliubov quasiparticle with energy $E_\alpha(\bm{p})\geq0$:
\bea
P_\alpha(\bm{p})&=&\Theta\big[B+E_\alpha(\bm{p})\big]-\Theta\big[B-E_\alpha(\bm{p})\big]\no\\
&\equiv&\Theta\big[E_\alpha(\bm{p})-|B|\big]\,,
\eea

\noi which only takes the values $P_\alpha(\bm{p})=\{0,1\}$ given that $E_\alpha(\bm{p})\neq|B|$. We observe that due to the chiral symmetry dictating the Hamiltonian $\hat{H}(\bm{p})$, the parity function is symmetric with respect to $B\leftrightarrow-B$ and changes from one to zero when the Zeeman energy crosses one the two chiral-symmetry-related levels $\pm E_\alpha(\bm{p})$. One also finds that in the special case of $\Delta_\alpha(\bm{p})=\Delta$ for all $\alpha$, the pa\-ri\-ties for all bands satisfy $P_\alpha(\bm{p})=1$ given that $\Delta>|B|$, hence reflecting that the SC remains fully gapped in spite of the presence of the Zeeman field.

At this stage, one separates intra- and inter-band contributions, cf Ref.~\onlinecite{Torma,PeottaLieb,LongLiang,PeottaTormaBAB,RossiRev}. For the intra-band $\alpha=\beta$ contribution, the matrix element $\big<u_\alpha(\bm{p})\big|\partial_{\bm{p}}\hat{h}(\bm{p})\big|u_\beta(\bm{p})\big>$ is simply given by $\partial_{\bm{p}}\varepsilon_\alpha(\bm{p})$. On the other hand, to infer the inter-band $\alpha\neq\beta$ contribution, we make use of the standard relation:
\begin{align}
\big<u_\alpha(\bm{p})\big|\partial_{\bm{p}}\hat{h}(\bm{p})\big|u_\beta(\bm{p})\big>=[\varepsilon_\beta(\bm{p})-\varepsilon_\alpha(\bm{p})]\big<u_\alpha(\bm{p})\big|\partial_{\bm{p}}u_\beta(\bm{p})\big>.\no
\end{align}

\noi The above considerations lead to the following result for the intra-band (also termed conventional) contribution:
\bea
D_{ij}^{\rm intra}&=&\sum_\alpha\int_{\rm BZ}\frac{d\bm{p}}{(2\pi)^d}\Big\{P_\alpha(\bm{p})-|B|\delta\big[E_\alpha(\bm{p})-|B|\big]\Big\}\no\\
&&\cdot\frac{\Delta_\alpha^2(\bm{p})}{E_\alpha(\bm{p})}\frac{\partial_{p_i}\varepsilon_\alpha(\bm{p})}{E_\alpha(\bm{p})}\frac{\partial_{p_j}\varepsilon_\alpha(\bm{p})}{E_\alpha(\bm{p})},\,\qquad
\label{eq:Dintra}
\eea

\noi as well as to the relation for the inter-band one:
\bea
&&D^{\rm inter}_{ij}=\sum_{\alpha\neq\beta}\int_{\rm BZ}\frac{d\bm{p}}{(2\pi)^d}\,\Delta_\alpha(\bm{p})P_\alpha(\bm{p})\frac{2[\varepsilon_\beta(\bm{p})-\varepsilon_\alpha(\bm{p})]^2}{E_\beta^2(\bm{p})-E_\alpha^2(\bm{p})}\no\\
&&\cdot\frac{\Delta_\beta(\bm{p})}{E_\alpha(\bm{p})}\big[\big<\partial_{p_i}u_\alpha(\bm{p})\big|u_\beta(\bm{p})\big>\big<u_\beta(\bm{p})\big|\partial_{p_j}u_\alpha(\bm{p})\big>+i\leftrightarrow j\big].\no\\
\label{eq:Dinter}
\eea

\noi Notably, since the factor $[\varepsilon_\beta(\bm{p})-\varepsilon_\alpha(\bm{p})]^2/[E_\beta^2(\bm{p})-E_\alpha^2(\bm{p})]$ goes to zero for $\alpha=\beta$, the constraint $\alpha\neq\beta$ can be lifted from the above expression.

This property allows us to express the inter-band contribution to the superfluid stiffness as a band-dependent sum. Specifically, after introducing the operator:
\begin{align}
\hat{M}_\alpha(\bm{p})=2\,\frac{\big[\hat{h}(\bm{p})-\varepsilon_\alpha(\bm{p})\big]^2}{\hat{E}^2(\bm{p})-E_\alpha^2(\bm{p})}\frac{\hat{\Delta}}{E_\alpha(\bm{p})}\end{align}

\noi we obtain the following compact expression:
\bea
D^{\rm inter}_{ij}&=&\sum_\alpha\int_{\rm BZ}\frac{d\bm{p}}{(2\pi)^d}\,\Delta_\alpha(\bm{p})P_\alpha(\bm{p})\no\\
&&\cdot\big[\big<\partial_{p_i}u_\alpha(\bm{p})\big|\hat{M}_\alpha(\bm{p})\big|\partial_{p_j}u_\alpha(\bm{p})\big>+i\leftrightarrow j\big].\qquad
\eea

Note that there may be cases in which there exist pairs of bands with $\alpha\neq\beta$ which satisfy $E_\alpha^2(\bm{p})-E_\beta^2(\bm{p})$ even though $\varepsilon_\alpha(\bm{p})\neq\varepsilon_\beta(\bm{p})$. In such situations, which take place for $\mu=0$, singularities may be introduced in $D_{ij}^{\rm inter}$. To avoid such issues, one can consider the evaluation of the inter-band superfluid stiffness tensor at $\mu=0$, by considering a nonzero $\mu$ and taking the limit $\mu\rightarrow0$. Finally, we note that in the special case where $\Delta_\alpha(\bm{p})=\Delta$ for all bands, the expressions for the $\hat{M}_\alpha(\bm{p})$ operator and the superfluid stiffness simplify according to:
\begin{align}
\hat{M}_\alpha(\bm{p})=2\,\frac{\hat{h}(\bm{p})-\varepsilon_\alpha(\bm{p})}{\hat{h}(\bm{p})+\varepsilon_\alpha(\bm{p})}\frac{\Delta}{E_\alpha(\bm{p})}\end{align}

\noi we obtain the following compact expression:
\bea
D^{\rm inter}_{ij}&=&\Delta\sum_\alpha\int_{\rm BZ}\frac{d\bm{p}}{(2\pi)^d}\,P_\alpha(\bm{p})\no\\
&&\cdot\big[\big<\partial_{p_i}u_\alpha(\bm{p})\big|\hat{M}_\alpha(\bm{p})\big|\partial_{p_j}u_\alpha(\bm{p})\big>+i\leftrightarrow j\big].\qquad
\eea

\section{Dyson Equation for a Spatially Varying Superconducting Phase}\label{app:AppendixC}

In the following paragraphs, we show how to describe the modified matrix Green function at first order in spatial gradients of the superconducting phase. We start from a coordinate space description. Since translational invariance is broken in the presence of the gradients, we need to define the single particle Green function at two positions $\bm{r}$ and $\bm{r}'$. We therefore obtain the expression for the modified Green function in the presence of a general local perturbation $\hat{V}(\bm{r})$:
\begin{align}
\hat{G}^{(1)}(\epsilon,\bm{r},\bm{r}')=\hat{G}(\epsilon,\bm{r}-\bm{r}')\qquad\qquad\qquad\qquad\phd\,\quad\no\\
+\int d\bar{\bm{r}}\ph
\hat{G}(\epsilon,\bm{r}-\bar{\bm{r}})\hat{V}(\bar{\bm{r}})
\hat{G}(\epsilon,\bar{\bm{r}}-\bm{r}').
\end{align}

\noi At this stage we express the matrix Green functions and the perturbation potential using Fourier transforms of the form $f(\bm{r})=\int \frac{d\bm{q}}{(2\pi)^d}\, e^{i\bm{q}\cdot\bm{r}}f(\bm{q})$ and end up with:
\bea
&&\hat{G}^{(1)}(\epsilon,\bm{r},\bm{r}')=\int\frac{d\bm{p}}{(2\pi)^d}\ph e^{i\bm{p}\cdot(\bm{r}-\bm{r}')}\hat{G}(\epsilon,\bm{p})\no\\
&&+\int\frac{d\bm{q}}{(2\pi)^d}\int\frac{d\bm{k}_1}{(2\pi)^d}\int d\bm{k}_2\ph \int\frac{d\bar{\bm{r}}}{(2\pi)^d}\ph e^{i(\bm{k}_2+\bm{q}-\bm{k}_1)\cdot\bar{\bm{r}}}\no\\
&&\qquad\quad\quad\times\ph
e^{i\bm{k}_1\cdot\bm{r}}e^{-i\bm{k}_2\cdot\bm{r}'}\hat{G}(\epsilon,\bm{k}_1)\hat{V}(\bm{q})
\hat{G}(\epsilon,\bm{k}_2).\qquad\quad
\eea

\noi Carrying out the integration over $\bar{\bm{r}}$ leads to the delta function $\delta(\bm{k}_2+\bm{q}-\bm{k}_1)$. To proceed, it is more convenient to write the resulting expression for the Green function in a symmetrized form, by splitting the second term in the above equation into two equal contributions stemming from scattering with wave vectors $\big(\bm{k}_1,\bm{k}_2\big)=\big(\bm{p}+\bm{q},\bm{p}\big)$ and $\big(\bm{k}_1,\bm{k}_2\big)=\big(\bm{p},\bm{p}-\bm{q}\big)$. This leads to the expression:
\begin{widetext}
\bea
\hat{G}^{(1)}\big(\epsilon,\bm{r},\bm{r}'\big)&=&\frac{1}{2}\int\frac{d\bm{p}}{(2\pi)^d}\ph e^{i\bm{p}\cdot(\bm{r}-\bm{r}')}\int \frac{d\bm{q}}{(2\pi)^d}\ph e^{i\bm{q}\cdot\bm{r}}\left[\hat{G}(\epsilon,\bm{p})(2\pi)^d\delta(\bm{q})+\hat{G}(\epsilon,\bm{p}+\bm{q})\hat{V}(\bm{q})\hat{G}(\epsilon,\bm{p})\right]\no\\
&+&\frac{1}{2}\int\frac{d\bm{p}}{(2\pi)^d}\ph e^{i\bm{p}\cdot(\bm{r}-\bm{r}')}\int \frac{d\bm{q}}{(2\pi)^d}\ph e^{i\bm{q}\cdot\bm{r}'}\left[\hat{G}(\epsilon,\bm{p})(2\pi)^d\delta(\bm{q})+\hat{G}(\epsilon,\bm{p})\hat{V}(\bm{q})\hat{G}(\epsilon,\bm{p}-\bm{q})\right]\,.\label{eq:modGreenCoordinate}
\eea
\end{widetext}

To proceed, it is more convenient to introduce the center of mass $\bm{R}=(\bm{r}+\bm{r}')/2$ and position difference $\bm{\delta}=\bm{r}-\bm{r}'$ coordinates. Since our primary goal is to obtain a translationally invariant modified Green function which is independent of $\bm{r}$ and $\bm{r}'$, we can approximately set $\bm{r},\bm{r}'\approx\bm{R}$, and obtain the matrix Green function in momentum space through the definition:
\begin{align}
\hat{G}^{(1)}(\epsilon,\bm{p},\bm{q})=
\int d\bm{p}\ph e^{-i\bm{p}\cdot\bm{\delta}}\int d\bm{q}\ph e^{-i\bm{q}\cdot\bm{R}}\ph\hat{G}^{(1)}\big(\epsilon,\bm{\delta},\bm{R}\big)\,,
\end{align}

\noi where $\hat{G}^{(1)}\big(\epsilon,\bm{\delta},\bm{R}\big)$ is found from Eq.~\eqref{eq:modGreenCoordinate} after repla\-cing $\bm{r}$ and $\bm{r}'$ by $\bm{\delta}$ and $\bm{R}$. By relying on the above, we exploit Eq.~\eqref{eq:modGreenCoordinate} and after replacing $\hat{V}(\bm{q})$ by $-\hat{\Delta}\phi(\bm{q})\tau_2$, we immediately obtain the momentum space defined modified matrix Green function $\hat{G}^{(1)}(\epsilon,\bm{p})$ in Eq.~\eqref{eq:modGreenMomentum}.

\section{Relation between Superfluid Stiffness and Synthetic Winding Number in 3D}\label{app:AppendixD}

In this appendix, we show some of the steps that allow us to go from Eq.~\eqref{eq:SSperm} to Eq.~\eqref{eq:SSfinal}. It is more convenient to demonstrate the equivalences of the two expressions by transferring to the frame in which the operator $\hat{\Pi}$ effecting chiral symmetry becomes block diagonal, i.e., it reads as $\hat{\Pi}={\rm diag}\{\mathds{1}_\tau,-\mathds{1}_\tau\}$. In the same basis, the adiabatic Hamiltonian takes the block off-diagonal form:
\begin{align}
\hat{\cal H}(\bm{p},\phi)=\left(\begin{array}{cc}\hat{0}&\hat{A}(\bm{p},\phi)\\\hat{A}^\dag(\bm{p},\phi)&\hat{0}\end{array}\right)\,.
\end{align}

\noi In the case of an adiabatic Hamiltonian which satisfies $\big[\hat{\cal H}(\bm{p},\phi)\big]^2=E^2(\bm{p})\mathds{1}$, one can define the normalized off-diagonal block Hamiltonian $\underline{\hat{A}}(\bm{p},\phi)=\hat{A}(\bm{p},\phi)/E(\bm{p})$, which satisfies the relation $\underline{\hat{A}}(\bm{p},\phi)\underline{\hat{A}}^\dag(\bm{p},\phi)=\mathds{1}_\sigma$. All the conduction and valence eigenstates are respectively degenerate with energies $\pm E(\bm{p})$, and their eigenvectors $\big|{\rm U}_\pm(\bm{p},\phi)\big>$ are given by the expression:
\begin{align}
\big|{\rm U}_\pm(\bm{p},\phi)\big>=\frac{1}{\sqrt{2}}\left(\begin{array}{c}\mathds{1}_\sigma\\\pm\underline{\hat{A}}^\dag(\bm{p},\phi)\end{array}\right)\,.
\end{align}

\noi In the basis of the conduction and valence subspaces, we have $\big<{\rm U}_{s}(\bm{p},\phi)\big|\hat{\Pi}\big|{\rm U}_{s'}(\bm{p},\phi)\big>=\frac{1-ss'}{2}\mathds{1}\equiv\delta_{s',-s}\mathds{1}$.

By introducing resolutions of identity in the conduction/valence band spaces in Eq.~\eqref{eq:SSperm}, we find that the superfluid stiffness can be re-expressed as:
\bea
D=-\frac{\varepsilon_{ijk}}{3}\int dP\sum_{s,s',s''}{\rm tr}\Big[\big<{\rm U}_{-s}\big|\partial_{\tilde{p}_i}\hat{\cal H}\big|{\rm U}_{s'}\big>\hat{\cal G}_{s'}\no\\
\big<{\rm U}_{s'}\big|\partial_{\tilde{p}_j}\hat{\cal H}\big|{\rm U}_{s''}\big>\hat{\cal G}_{s''}\big<{\rm U}_{s''}\big|\partial_{\tilde{p}_k}\hat{\cal H}\big|{\rm U}_{s}\big>\hat{\cal G}_{s}\Big]\label{eq:SStobesimplified}
\eea

\noi where for convenience we suppressed the arguments $(\bm{p},\phi)\equiv(\tilde{p}_1,\tilde{p}_2,\tilde{p}_3)$ and introduced the matrix Green functions $\hat{\cal G}_{s}$ which are projected onto the conduction/valence subspace and read as:
\begin{align}
\hat{\cal G}_{s}\equiv\big<{\rm U}_s\big|\hat{\cal G}\big|{\rm U}_s\big>=\frac{\mathds{1}_\sigma}{i\epsilon+B-sE(\bm{p})}\equiv{\cal G}_s\mathds{1}_\sigma\,,
\end{align}

\noi where ${\cal G}_s(\bm{p},\phi)=1/\big[i\epsilon+B-sE(\bm{p})\big]$. The above form results from the degeneracy of the states within the conduction (similarly for the valence) subspace.

To proceed, we now obtain concrete expressions for the matrix elements involving derivatives of the Hamiltonian. We find the following expression for elements involving states of a given conduction/valence subspace:
\bea
\big<{\rm U}_s\big|\partial_{\tilde{p}_i}\hat{\cal H}\big|{\rm U}_s\big>=s\partial_{\tilde{p}_i}E\mathds{1}_\sigma\,.
\eea

\noi In contrast when such a derivative  involves one state from the conduction band and one from the valence band, we find the result:
\bea
\big<{\rm U}_s\big|\partial_{\tilde{p}_i}\hat{\cal H}\big|{\rm U}_{-s}\big>=sE\underline{\hat{A}}\partial_{\tilde{p}_i}\underline{\hat{A}}^\dag\,.
\eea

Plugging the above results into Eq.~\eqref{eq:SStobesimplified}, the antisymmetric tensor $\varepsilon_{ijk}$ implies that the superfluid stiffness is obtained only by the following contribution:
\bea
D&=&\frac{\varepsilon_{ijk}}{3}\int \frac{d\bm{p}}{(2\pi)^2}\,E^3{\rm tr}\Big[\Big(\underline{\hat{A}}\partial_{\tilde{p}_i}\underline{\hat{A}}^\dag\Big)\Big(\underline{\hat{A}}\partial_{\tilde{p}_j}\underline{\hat{A}}^\dag\Big)\Big(\underline{\hat{A}}\partial_{\tilde{p}_k}\underline{\hat{A}}^\dag\Big)\Big]\no\\
&&\cdot\int_{-\infty}^{+\infty}\frac{d\epsilon}{2\pi}\sum_{s=\pm}s\,{\cal G}_s{\cal G}_{-s}^2\,.\label{eq:SSintermediate}
\eea

\noi The evaluation of the integral in the second row yields:
\begin{align}
\int_{-\infty}^{+\infty}\frac{d\epsilon}{2\pi}\sum_{s=\pm}s\,{\cal G}_s{\cal G}_{-s}^2=\frac{1}{2}\frac{d}{dE}\left[\frac{\Theta\big(E-|B|\big)}{E}\right]\,.\label{eq:GrIntegral}
\end{align}

To demonstrate the equivalences of Eqs.~\eqref{eq:SSperm} and~\eqref{eq:SSfinal} it is required to relate the term:
\begin{align}
\varepsilon_{ijk}\Big[\Big(\underline{\hat{A}}\partial_{\tilde{p}_i}\underline{\hat{A}}^\dag\Big)\Big(\underline{\hat{A}}\partial_{\tilde{p}_j}\underline{\hat{A}}^\dag\Big)\Big(\underline{\hat{A}}\partial_{\tilde{p}_k}\underline{\hat{A}}^\dag\Big)\Big]\no
\end{align}

\noi to the winding number density defined in Eq.~\eqref{eq:winding3densitymomentum}. Alternatively, we start from Eq.~\eqref{eq:winding3densitymomentum} and express it in terms of the above term. By employing the basis in which the adiabatic Hamiltonian is block off-diagonal, we find that:
\begin{align}
\hat{\cal H}^{-1}(\bm{p},\phi)=\left(\begin{array}{cc}\hat{0}&\big[\hat{A}^\dag(\bm{p},\phi)\big]^{-1}\\\hat{A}^{-1}(\bm{p},\phi)&\hat{0}\end{array}\right)\,.
\end{align}

\noi The above implies that the winding number density simplifies to:
\begin{align}
w_3(\bm{p},\phi)=-\frac{\varepsilon_{ijk}}{3!}{\rm tr}\Bigg\{\left(\hat{A}^{-1}\partial_{\tilde{p}_i}\hat{A}\right)\left(\hat{A}^{-1}\partial_{\tilde{p}_j}\hat{A}\right)\left(\hat{A}^{-1}\partial_{\tilde{p}_k}\hat{A}\right)\no\\
-\left[\Big(\hat{A}^\dag\Big)^{-1}\partial_{\tilde{p}_i}\hat{A}^\dag\right]\left[\Big(\hat{A}^\dag\Big)^{-1}\partial_{\tilde{p}_j}\hat{A}^\dag\right]\left[\Big(\hat{A}^\dag\Big)^{-1}\partial_{\tilde{p}_k}\hat{A}^\dag\right]\Bigg\}/2.\no
\end{align}

\noi By virtue of the fact that the winding density is a real number, we can take the hermitian conjugate of the term in the second row in the expression above, and find that:
\begin{align}
w_3(\bm{p},\phi)=-\frac{\varepsilon_{ijk}}{3!}{\rm tr}\Bigg[\left(\hat{A}^{-1}\partial_{\tilde{p}_i}\hat{A}\right)\left(\hat{A}^{-1}\partial_{\tilde{p}_j}\hat{A}\right)\left(\hat{A}^{-1}\partial_{\tilde{p}_k}\hat{A}\right)\Bigg]\no.
\end{align}

\noi where we made use of the relation $\Big[\big(\hat{A}^\dag\big)^{-1}\Big]^\dag=\hat{A}^{-1}$. By further taking into account that $\hat{A}^{-1}=\hat{A}^\dag/E^2=\underline{\hat{A}}^\dag/E$ we write:
\begin{align}
w_3(\bm{p},\phi)=-\frac{\varepsilon_{ijk}}{3!E^3}{\rm tr}\Bigg[\left(\underline{\hat{A}}^\dag\partial_{\tilde{p}_i}\hat{A}\right)\left(\underline{\hat{A}}^\dag\partial_{\tilde{p}_j}\hat{A}\right)\left(\underline{\hat{A}}^\dag\partial_{\tilde{p}_k}\hat{A}\right)\Bigg]\no.
\end{align}

\noi Moreover, by virtue of the combined presence of the antisymmetric symbol $\varepsilon_{ijk}$ and the ${\rm tr}$ operation, the above result can be solely rewritten in terms of $\underline{\hat{A}}$, and we find:
\begin{align}
w_3(\bm{p},\phi)=\frac{\varepsilon_{ijk}}{3!}{\rm tr}\Bigg[\left(\underline{\hat{A}}\partial_{\tilde{p}_i}\underline{\hat{A}}^\dag\right)\left(\underline{\hat{A}}\partial_{\tilde{p}_j}\underline{\hat{A}}^\dag\right)\left(\underline{\hat{A}}\partial_{\tilde{p}_k}\underline{\hat{A}}^\dag\right)\Bigg]
\label{eq:windingdensityBlockOff}
\end{align}

\noi where we also took into account that $\underline{\hat{A}}\partial\underline{\hat{A}}^\dag=-\underline{\hat{A}}^\dag\partial\underline{\hat{A}}$ and made use of the cyclic property of the trace. By direct comparison, we find that Eq.~\eqref{eq:SSintermediate} in conjunction with Eqs.~\eqref{eq:GrIntegral} and~\eqref{eq:windingdensityBlockOff}, lead to the equivalence of Eqs.~\eqref{eq:SSperm} and~\eqref{eq:SSfinal}.

{\color{black}
\section{Superfluid Stiffness of Higher-Order Band Touching Points - Adiabatic Formalism}\label{app:AppendixE}

In this section, we show that our adiabatic for\-ma\-lism is naturally capable of also recovering the result of Eq.~\eqref{eq:HOTBTPstiffness} obtained for the Hamiltonian of Eq.~\eqref{eq:HOTBTP} which describes a single higher-order BTP. For this purpose, we introduce the nonzero components:
\bea
d_x(\bm{p})&=&\varepsilon_D\left(\frac{p}{p_D}\right)^{|s|}\cos\big[s\theta(\bm{p})\big]\,,\\
d_y(\bm{p})&=&\varepsilon_D\left(\frac{p}{p_D}\right)^{|s|}\sin\big[s\theta(\bm{p})\big]\,,
\eea

\noi of the vector $\bm{d}(\bm{p})$ which parametrizes the Hamiltonian of Eq.~\eqref{eq:HOTBTP} according to $\hat{h}^{(s)}(\bm{p})=\bm{d}(\bm{p})\cdot\bm{\sigma}$.

We now start from the fundamental expression of the superfluid stiffness in Eq.~\eqref{eq:StiffnessDirac2Dw3} and use the derivative chain rule to write:
\bea
D^{(s)}=\sum_i^{x,y}\int dP\ph{\rm Tr}\left[\frac{\partial\hat{h}^{(s)}(\bm{p})}{\partial p_i}\mathds{1}_\tau\hat{\cal F}_{p_i\phi}(\epsilon,\bm{p},\phi)\right]\qquad\qquad\phd\,\no\\
=\sum_{i,j,k}^{x,y}\int dP\ph\frac{\partial d_j(\bm{p})}{dp_i}\frac{\partial d_k(\bm{p})}{dp_i}\ph{\rm Tr}\left[\frac{\partial\hat{h}^{(s)}(\bm{d})}{\partial d_j}\mathds{1}_\tau\hat{\cal F}_{d_k\phi}(\epsilon,\bm{d},\phi)\right]\no\\
=\sum_{i,j}^{x,y}\int dP\ph\left[\frac{\partial d_j(\bm{p})}{dp_i}\right]^2\ph{\rm Tr}\left[\frac{\partial\hat{h}^{(s)}(\bm{d})}{\partial d_j}\mathds{1}_\tau\hat{\cal F}_{d_j\phi}(\epsilon,\bm{d},\phi)\right],\,\,\,\,\,\,\,\no
\eea

\noi where we obtained the above result by observing that the trace in the second row is nonzero only when $j=k$ for $\mu=0$. To proceed, we make use of the relations below:
\bea
\frac{\partial d_x(\bm{p})}{\partial p_x}=+\frac{\partial d_y(\bm{p})}{\partial p_y}&=&+
|s|\frac{\varepsilon(p)}{p}\cos\big[\big(|s|-1\big)\theta(\bm{p})\big],\no\\
\frac{\partial d_x(\bm{p})}{\partial p_y}=-\frac{\partial d_y(\bm{p})}{\partial p_x}&=&-
|s|\frac{\varepsilon(p)}{p}\sin\big[\big(|s|-1\big)\theta(\bm{p})\big],\no
\eea

\noi where we introduced the modulus $\varepsilon(\bm{p})=|\bm{d}(\bm{p})|$, which here only depends on the modulus of the momentum, i.e., $\varepsilon(\bm{p})\equiv \varepsilon(p)=\varepsilon_D\big(p/p_D\big)^{|s|}$. From the above, we find:
\begin{align}
\sum_i^{x,y} \left[\frac{\partial d_x(\bm{p})}{dp_i}\right]^2=\sum_i^{x,y} \left[\frac{\partial d_y(\bm{p})}{dp_i}\right]^2=\left[|s|\frac{\varepsilon(p)}{p}\right]^2\,,
\end{align}

\noi which subsequently leads to:
\begin{align}
D^{(s)}=
\sum_i^{x,y}\int dP\left[|s|\frac{\varepsilon(p)}{p}\right]^2{\rm Tr}\left[\frac{\partial\hat{h}^{(s)}(\bm{d})}{\partial d_i}\mathds{1}_\tau\hat{\cal F}_{d_i\phi}(\epsilon,\bm{d},\phi)\right].\no
\end{align}

\noi We note that the contribution of the trace is a function of only $p$ or equivalently $\varepsilon$. Hence, by denoting this $f(p)$ and $f(\varepsilon)$, respectively, we make use of the following relations:
\bea
&&\int \frac{d\bm{p}}{\big(2\pi\big)^2}\left[|s|\frac{\varepsilon(p)}{p}\right]^2f(p)=\int_0^\infty \frac{dp\, p}{2\pi}\left[|s|\frac{\varepsilon(p)}{p}\right]^2f(p)\no\\
&&=|s|\int_0^\infty\frac{dp}{2\pi}\frac{d\varepsilon(p)}{dp}\varepsilon(p)f(p)\equiv|s|\int\frac{d\bm{d}}{\big(2\pi\big)^2}\ph f(\varepsilon)\,.
\eea

\noi From the above, we obtain that the superfluid stiffness can be written in terms of the two components of $\bm{d}$ as:
\bea
D^{(s)}=|s|\sum_i^{x,y}\int d{\cal D}\ph{\rm Tr}\left[\frac{\partial\hat{h}^{(s)}(\bm{d})}{\partial d_i}\mathds{1}_\tau\hat{\cal F}_{d_i\phi}(\epsilon,\bm{d},\phi)\right],\quad
\eea

\noi where we introduced the shorthand notation:
\begin{align}
\int d{\cal D}\equiv\int_{\rm BZ}\dfrac{d\bm{d}}{(2\pi)^d}\int_{-\infty}^{+\infty}\dfrac{d\epsilon}{2\pi}\,.\no
\end{align}

\noi Since, however, the vector $\bm{d}$ plays here an analogous role to $\upsilon_D\bm{p}$, the above expression implies that $
D^{(s)}(B,\mu=0)=|s|D^{(1)}(B,\mu=0)$. Hence, a BTP with vorticity $s$ yields a stiffness which is $|s|$ times that of a single Dirac point with vorticity of a single unit ($|s|=1$). Even more, one finds that the above relation actually holds for all $\mu$. This is more convenient to demonstrate by employing Eqs.~\eqref{eq:Dintra} and~\eqref{eq:Dinter}, after setting $\varepsilon_\alpha(\bm{p})=\alpha\varepsilon(p)-\mu$, $\Delta_\alpha(\bm{p})=\Delta$, and $\alpha=\pm$. Thus, we this provides that:
\begin{align}
D^{(s)}(B,\mu)=|s|D^{(1)}(B,\mu)\,.
\end{align}

We note that $D^{(1)}(B=0,\mu)$ was first obtained in Ref.~\onlinecite{KopninSoninPRB}. Later on, it was shown in Ref.~\onlinecite{LongLiang} that it consists of the intra- and inter-band contributions given by $D^{(1)}_{\rm intra}(B=0,\mu)=\big(|\mu|/2\pi\big)\sqrt{1+v^2}$ and $D^{(1)}_{\rm inter}(B=0,\mu)=\big(|\mu|/2\pi\big)v^2\ln\big[\big(1+\sqrt{1+v^2}\big)/v\big]$, respectively, where $v=\Delta/|\mu|$. Note that in the limit $v\rightarrow0$ only the intra-band contribution survives and yields the result $D^{(1)}(B=0,\mu;\Delta\ll|\mu|)=|\mu|/2\pi$. In this limit, only one out of the two $\pm\upsilon_D p$ bands crosses the Fermi energy which is determined by the value of $|\mu|$. The noncrossing band lies energetically far away and does not contribute to the stiffness.
}

\end{document}